\documentclass[%
 reprint,superscriptaddress,
 amsmath,amssymb,
 aps,
]{revtex4-2}

\usepackage{graphicx}

\usepackage{dsfont}
\usepackage{dcolumn}
\usepackage{bm}
\usepackage[mathlines]{lineno}

\usepackage{diagbox}   
\usepackage{booktabs}  
\usepackage{array}     

\usepackage{titlesec}

\newcommand{\mainSectionFormat}{
  \titleformat{\section}[runin]
    {\bfseries}{}
    {0pt}{}[.—]
  \titlespacing*{\section}
    {0pt}{0pt}{0.5em}
}

\mainSectionFormat  
  
\begin{document}
\mainSectionFormat

\title{Bound States in Second-order Topological Graphitic Structures}

\author{Haiyue Huang}
\affiliation{Division of Physical Sciences, College of Letters and Science, University of California, Los Angeles, 90095, California, USA}

\author{Prineha Narang}
\thanks{Contact authors}
\affiliation{Division of Physical Sciences, College of Letters and Science, University of California, Los Angeles, 90095, California, USA}
\affiliation{Department of Electrical and Computer Engineering, University of California, Los Angeles, 90095, California, USA}

\author{Ioannis Petrides}
\thanks{Contact authors}
\affiliation{Division of Physical Sciences, College of Letters and Science, University of California, Los Angeles, 90095, California, USA}

\begin{abstract}
Quadrupole insulators are a class of second-order topological insulators (SOTIs) that host zero-dimensional corner states within a two-dimensional bulk. 
Despite their unique properties, their realization in electronic systems on realistic material platforms remains rare. 
In this work, we present a general design principle to obtain quadrupole insulators based on two-dimensional graphitic structures.
By engineering the positions and connections of zigzag edges, we identify four topological classes of graphitic structures.
We show that topologically protected massless corner state emerge at the intersection of domains belonging to different topological classes. 
Crucially, by tuning the smoothness of the domain wall, we further demonstrate the appearance of additional massive localized states with non-zero angular momentum.  
Our results provide a practical framework for realizing experimentally accessible SOTIs and uncover the coexistence of both massless and massive bound states in two dimensions.

\end{abstract}

\maketitle


\section{\label{sec:intro} Introduction}
According to the bulk-boundary correspondence, gapless boundary states are one of the manifestations of nontrivial topological insulators (TIs)~\cite{kane_z_2_2005,volkov1985two}.
Such states appear at the interface of two insulators that belong to distinct topological phases, with their energy and localization protected by the bulk topological index. 
Beyond gapless states, studies in one-dimensional (1D) systems have shown the possible existence of additional localized states, known as Volkov-Pankratov (VP) states \cite{tchoumakov2017volkov,lu2020dirac, van_den_berg_volkov-pankratov_2020, mahler2021massive}. 
In contrast to topologically protected gapless states, VP states are massive with energy depending on the characteristic profile of the domain wall that forms the interface. 
With the development of higher-order topological insulators, the bulk-boundary correspondence can be further generalized to higher co-dimension bound states~\cite{benalcazar_quantized_2017, benalcazar_electric_2017, schindler2018higher, petrides2020higher,petrides2022semiclassical, langbehn_reflection-symmetric_2017, xie_higher-order_2021, trifunovic_higher-order_2019, geier2018second}. 
For example, two-dimensional second-order topological insulators (2D SOTIs) host zero-dimensional (0D) localized corner states, as opposed to 1D boundary states in conventional topological insulators. In higher dimensions, however, the interplay between domain walls and the full spectrum of both massless and massive corner states has yet to be explored. 


To fill this gap, a material platform with an adjustable domain wall is required. Several material candidates have been proposed for 2D SOTIs \cite{park_higher-order_2019, hu_intrinsic_2022, ni2022organic, sheng_two-dimensional_2019, liu_two-dimensional_2019, ezawa2018minimal, PhysRevB.111.035113, radha_buckled_2020, pan2022two, kempkes_robust_2019, xue_higher-order_2021, PhysRevB.104.245427, mu2022kekule, PhysRevB.106.085126, PhysRevLett.126.066401, hu2023identifying}. 
In those cases, however, corner states exist at a material/vacuum boundary. 
In other words, the domain wall is defined by the finite size of the sample itself. This geometric constraint makes it almost impossible to study the effects of the domain wall and the emergence of massive bound states. 

Quadrupole insulators \cite{benalcazar_quantized_2017} are among the first proposed SOTIs, and the existence of four distinct topological classes within their framework provides a practical way to engineer domain walls of varying smoothness. 
While experimental realizations have been reported across various platforms \cite{he2020quadrupole, xie_higher-order_2021, peterson2018quantized, mittal2019photonic, imhof2018topolectrical, serra2018observation, yamada2022bound}, an electronic realization of quadrupole insulators with all four topological classes remains to be demonstrated.
In this work, we propose a systematic way to construct 2D SOTIs using graphitic structures, more specifically quadrupole topological insulators. 
We show a complete phase diagram comprising four representative structures, each belonging to a unique topological class.  With the four classes of structures in hand, we then connect them together with a steep or a smooth domain wall. Our results demonstrate that, in addition to massless corner states, massive localized states emerge when a smooth domain wall is employed.


\begin{figure}
\includegraphics[width=\columnwidth]{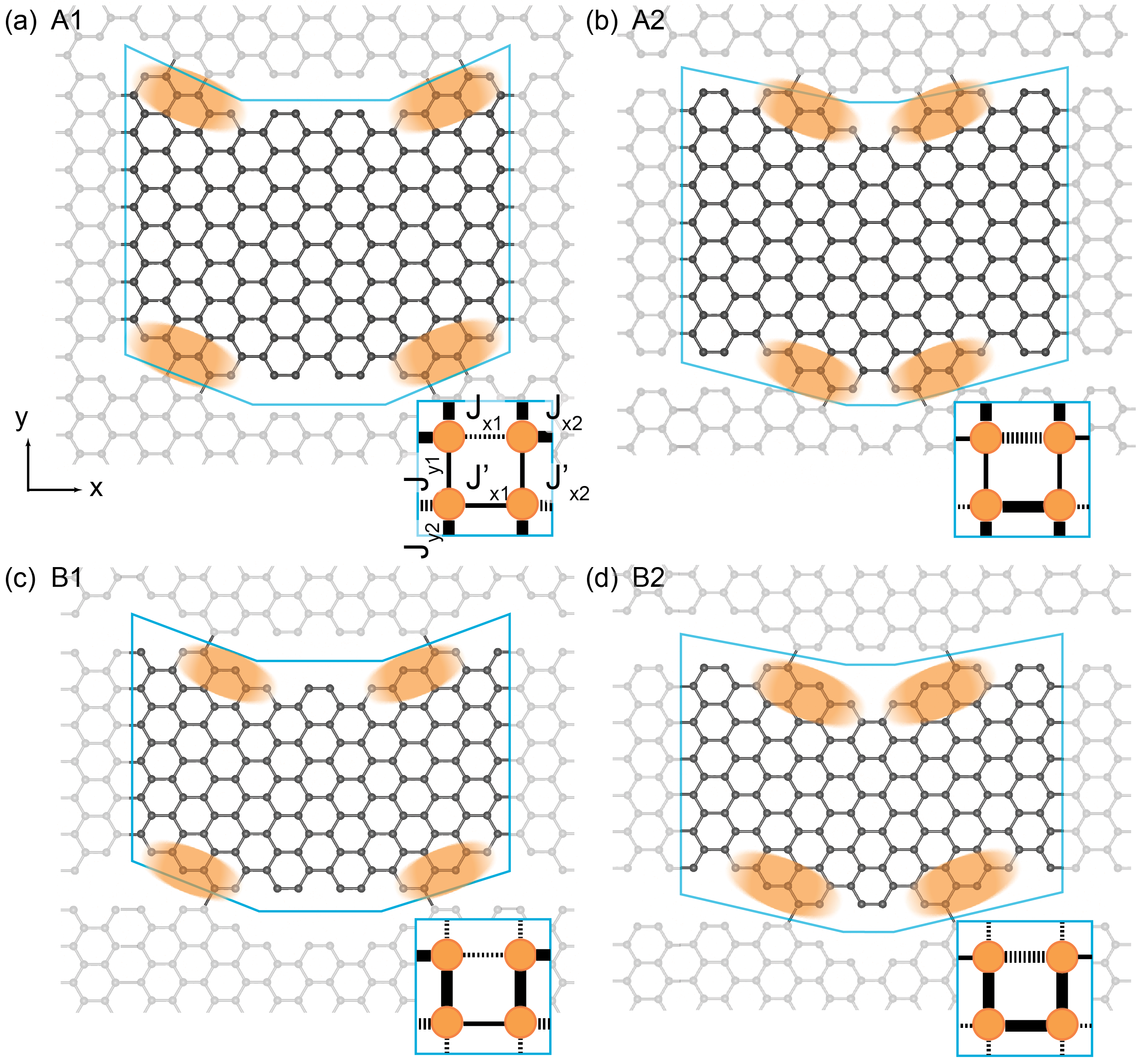}
\caption{\label{fig:schematic} The relative strength of intra-cell ($J_{\mu 1}$) and inter-cell ($J_{\mu 2}$) hopping along the $x$ and $y$ directions give rise to four topological classes of graphitic structures. (a--d) are representative unit cells made of conjugated carbon atoms belonging to each class, with corresponding simplified tight-binding models illustrated in the insets. Orange shades mark zigzag edges and blue lines are boundaries of unit cells. Solid and dashed lines indicate positive (negative) sign of the hopping parameter. Thick (thin) lines indicate a larger (smaller) hopping strength. Carbon atoms on the edge are saturated with hydrogen atoms, which are not displayed.}
\end{figure}

\section{\label{sec:sec1} Theory}

We consider a 2D Hamiltonian,
\begin{align}
 \hat{h}=& \sum_{{\mathbf{m}}}\left[\hat{H}_{x}(\mathbf{m} )+\hat{H}_{y}(\mathbf{m}) \right]\,,
\label{HamProdK}
\end{align}
where ${\mathbf{m}}=(m_x,m_y)$ are the lattice vectors,  
\begin{align}
\hat{H}_{x}({\mathbf{m}}  )=& J_{x1} c^\dagger_{{\mathbf{m}}+\mathbf{e}_x}c^{}_{{\mathbf{m}} } +J_{x2}  c^\dagger_{{\mathbf{m}}-\mathbf{e}_x } c^{}_{{\mathbf{m}}  } \,,\label{eq:Creutz}\\
\hat{H} _{y} ({\mathbf{m}} )=&e^{im_x \pi} J_{y1} c^\dagger_{{\mathbf{m}}+\mathbf{e}_y }c^{}_{{\mathbf{m}}  } +e^{-im_x \pi} J_{y2}  c^\dagger_{{\mathbf{m}}-\mathbf{e}_y   } c^{}_{{\mathbf{m}}  }\,,
\label{eq:SSH}
\end{align}
and $J_{\mu1}$ ($J_{\mu2}$) is the intra-cell (inter-cell) hopping strength.

The resulting lattice is made out of Su-Schrieffer-Heeger (SSH) chains $\hat{H}_{\mu}({\mathbf{m}}  )$ in both the $x$- and $y$-directions, and each $xy$-plaquette is threaded by a magnetic field with $\pi$ flux quanta. 
As a function of $J_{\mu}$, the spectrum of $\hat{h}$ has: (i) 1D edge states appearing below zero energy that merge into the bulk, and (ii) zero-energy 0D corner states that merge into the edge or bulk spectrum at the gap closing points.

Linearizing the dynamics of $\hat{h}$ around zero energy we find that the low-energy theory is given by a single Dirac cone around $\textbf{k} \sim (\pi,\pi)$, described by the Hamiltonian 
\begin{eqnarray}
\hat{H}_{0} = \mathbf{{d}}_0\cdot\mathbf{\Gamma}\,,
\label{eq:Model I}
\end{eqnarray}
where $\mathbf{d}_0=\{\mu_0 , \mu_x , \mu_y, {\nu}_x \mathrm{k}_x, {\nu}_y  \mathrm{k}_y  \}$, $\nu_\mu = J_{\mu2}  $, $\mu_\mu = J_{\mu2} -J_{\mu1} $, $\mu_0 $ is a chiral-breaking mass, and $\mathbf{\Gamma} = \{ \Gamma_0, \Gamma_1 ,\Gamma_2 ,\Gamma_3 , \Gamma_4 \}$ are five matrices,
\begin{equation}	
\begin{array}{c}	\Gamma_0 =\begin{pmatrix}     
		-\mathds{1} & 0 \\
		0 & \mathds{1} 
	\end{pmatrix}\,,
	\Gamma_1 =\begin{pmatrix}  
		0 &  \sigma_z \\
		\sigma_z & 0 
	\end{pmatrix}\,,
	\Gamma_2 =\begin{pmatrix}    
		0 & \sigma_x\\
		\sigma_x & 0 
	\end{pmatrix}\,,\\
	\Gamma_3 =\begin{pmatrix}    
		0 &- i\mathds{1} \\
		i\mathds{1} & 0  
	\end{pmatrix}\,,
	\Gamma_4 =\begin{pmatrix}     
		0 & \sigma_y\\
		\sigma_y & 0   
	\end{pmatrix}\,, 
\end{array}
\label{eq:basis I}
\end{equation}    
defining the Clifford algebra $\{\Gamma_i , \Gamma_j\} = 2\delta_{ij}\mathds{1}$, with $\{\sigma_x,\sigma_y,\sigma_z\}$ the Pauli matrices.  
The above model is equivalent to a free two-dimensional Dirac particle with linear dispersion $\nu_x$ ($\nu_y$) in the $x$- ($y$-) direction, in a background potential $\bm{\mu} = \{\mu_0,\mu_x,\mu_y\}$. 
For simplicity we take $\nu_x=\nu_y=\nu$.

The localized states around zero energy are found by solving the continuum Hamiltonian $\hat{H}_c=\mathbf{{d}}_c\cdot\mathbf{\Gamma}$, where ${\textbf{d}}_c = \{\mu_0 , \mu_x , \mu_y , -{\nu} i\partial_x, -{\nu}  i\partial_y \}$, with a plane wave ansatz $\psi = e^{i (\text{k}_x x + \text{k}_y y)}\phi$, where $\phi$ are the Bloch eigenvectors of $\hat{H}_0$. 
Since we are interested in finding co-dimension 2 states, i.e., states that are localized in both dimensions, we assume a topological domain wall formed by choosing
\begin{eqnarray}
    \mu_i = \begin{cases}
         -\Delta_i& x_i<-\lambda_i\\
         \frac{\Delta_i}{\lambda_i}x_i& -\lambda_i\leq x_i\leq \lambda_i \\
         \Delta_i & x_i> \lambda_i\end{cases}
\end{eqnarray}
where $2|\Delta_i|$ is the height of the domain wall and $\lambda_i$ its spread. 
For simplicity we choose $\Delta_x = \Delta_y=\Delta$ and $\lambda_x=\lambda_y=\lambda$


In order to calculate the localized states we square the Hamiltonian to obtain
\begin{eqnarray}
    \hat{H}_c^2 = -\nu^2\sum\limits_i  \partial^2_i + V(x,y)
    \label{eq:diff}
\end{eqnarray}
with $V(x,y)=\mu_0^2+\mu_x^2+\mu_y^2+\nu \partial_x \mu_x \Gamma_x+\nu \partial_y \mu_y\Gamma_y$, where $\Gamma_x=\sigma_z\otimes\sigma_z$, and $\Gamma_y=\mathds{1}\otimes\sigma_z$.
We first solve Eq.~\eqref{eq:diff} in the region $-\lambda_i\leq x_i\leq \lambda_i$ by changing to polar coordinates and separating variables such that $H^2\psi(r,\theta)=\epsilon^2 \psi(r,\theta)$ with $\psi(r,\theta) = Y(\theta) R(r)$.
From the differential equation, the angular part satisfies $\partial^2_\theta Y(\theta) =- l^2 Y(\theta)$, leading to $Y(\theta) = e^{il\theta}$, where $l$ is an integer describing the angular momentum of the state.
The radial part satisfies
\begin{eqnarray}
    \nu^2 R''(r) +\frac{\nu^2}{r}R'(r) -\left(\mu_0^2 + 2 \nu\Lambda \Sigma-\epsilon^2 + \Lambda^2r^2 + \nu^2\frac{l^2}{r^2}\right)R(r) = 0
    \label{eq:radial eq}
\end{eqnarray}
where $R(r)=\left[R_1(r),R_2(r),R_3(r),R_4(r)\right]$ is understood to be a 4-vector, $\Sigma = \text{diag}(1,-1,0,0)$ is a diagonal matrix, and $\Lambda=\Delta/\lambda$. 

The energies associated to each unit cell site are given by, see Sec. II of Supplemental Material, 
\begin{eqnarray}
    \epsilon_i ^{n,l}=\sqrt{\mu_0^2 +2\Lambda \nu(1+|l|+2n +\Sigma_{ii})}
    \label{eq:energy}
\end{eqnarray}

\begin{figure} 
\includegraphics[width=\columnwidth]{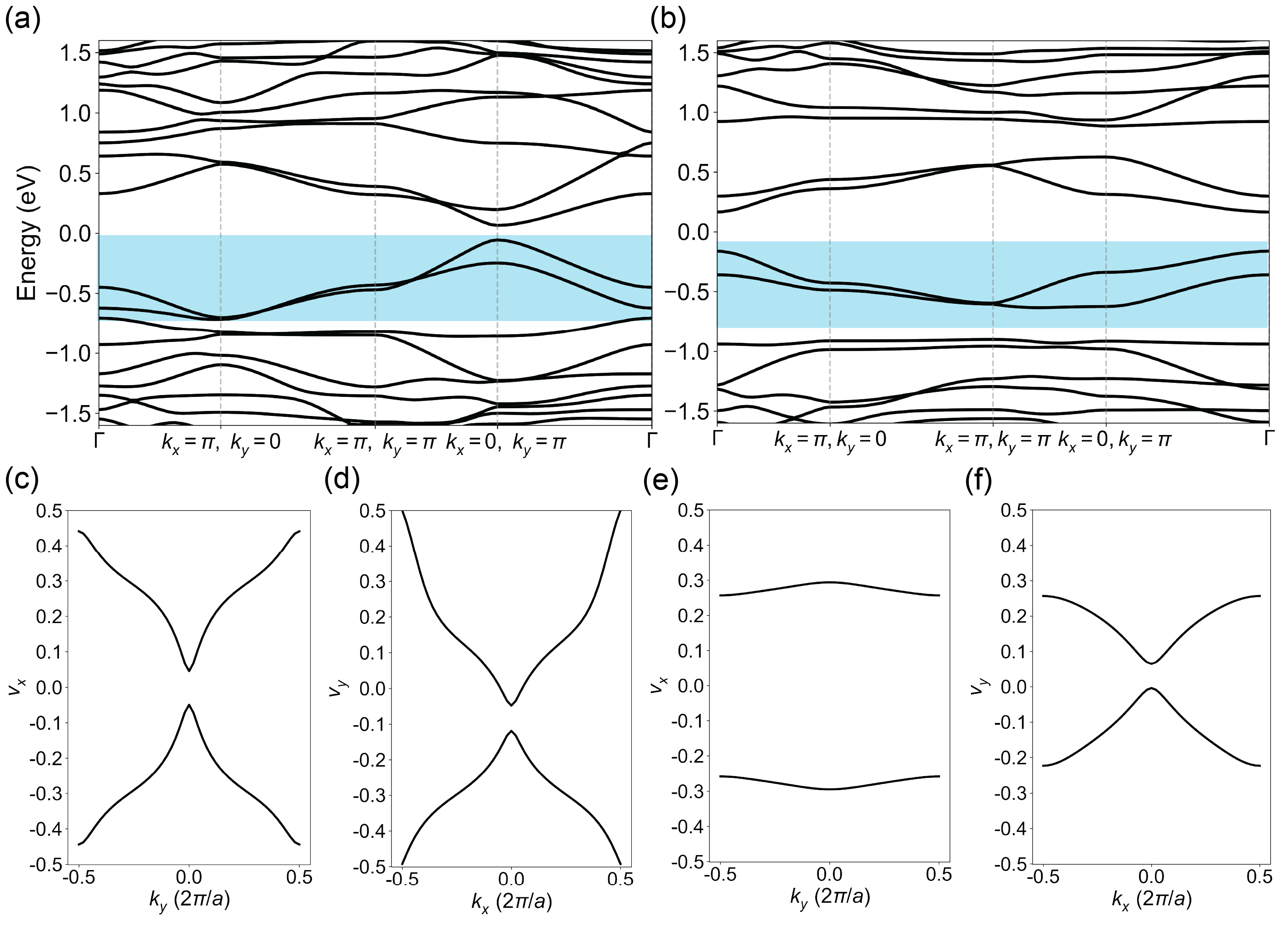}
\caption{\label{fig:periodic} Electronic band structures calculated with DFT method of (a) type A and (b) type B graphitic unit cells show middle bands separated from low-/high-lying bands. By calculating Wilson loop along the $x$ and $y$ directions on the two occupied middle bands (blue shaded), two Wannier bands are obtained. (c, d) and (e, f) are Wannier bands for the unit cell A1 and B1, respectively. Note there is a small gap in (d) at $k=\pi/a$. Energy values are given relative to the Fermi level.}
\end{figure}

\begin{figure}
\includegraphics[width=\columnwidth]{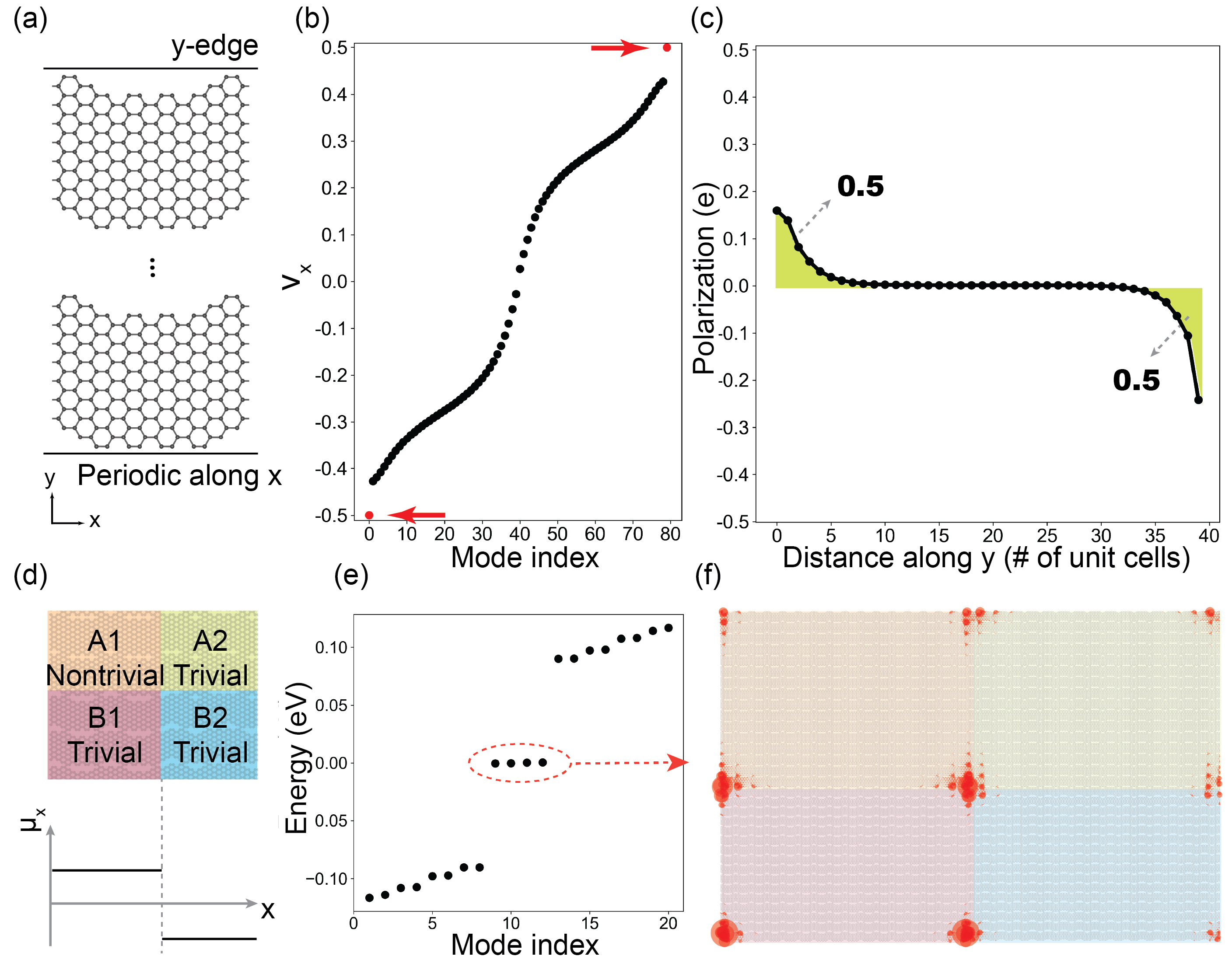}
\caption{\label{fig:thin} (a) Periodic $x$-ribbon structure made of $N_y=40$ repeating unit cells with open boundary condition along $y$ for unit cell A1. (b) By calculating Wilson loop, their Wannier centers for the $2N_y$ bands below Fermi level are obtained, showing pairs of edge states (red arrows). Due to the existence of $M_x$, edge states are degenerate with quantized $v_x=0$ or $0.5$. (c) Tangential polarization of the ribbon plot against distance along $y$, where the integrated polarization is 0.5$|e|$. Same plots for other unit cells can be found in Supplemental Material. (d) Supercell structure consisting of four distinct regions, each containing 16 $\times$ 16 graphitic unit cells. Regions corresponding to different topological classes are indicated by different colors. (e) Discrete energy levels near Fermi level, showing four degenerate in-gap states. (f) Charge distribution of the four in-gap states, showing localization at the corners of regions. }
\end{figure}

\section{Bulk Hamiltonian}
In Eq.~\ref{HamProdK}, the presence of hopping terms with negative signs is essential to ensure the nontriviality \cite{benalcazar_quantized_2017, petrides2020higher}. Here, we demonstrate that graphitic structures can be designed to inherently satisfy the requirement and thereby host a nontrivial higher-order topological phase.

Zigzag edges in graphene are known to host localized states with nearly zero energy \cite{ezawa2006peculiar, wakabayashi2010electronic}. 
The band structures and low-energy properties of zigzag terminated graphitic materials are thus dominated by the coupling of zigzag edges. 
This provides a versatile platform for designing materials that can be effectively described using a much simpler model \cite{huang2025topological, groning2018engineering, rizzo2018topological, rizzo2020inducing, cao_topological_2017}. 
Specifically, as shown in Fig.~\ref{fig:schematic}, zigzag edges (orange shades) are coupled along both $x$ and $y$ directions leading to an effective low-energy Hamiltonian given by the nearest-neighbor hopping of zigzag edges, as illustrated in the insets of Fig.~\ref{fig:schematic}. 
The zigzag edges adopt either a ``face-to-face" or a ``back-to-back" configuration, which leads to a positive or negative effective hopping. This correspondence between the configuration and the hopping sign has been observed and discussed in previous studies \cite{rizzo2018topological,joost2019correlated}. By varying the number of benzene columns between zigzag edges, we are able to control the relative magnitude of intra-cell ($J_{\mu1}$) and inter-cell ($J_{\mu2}$) hopping. 
The only exception is the inter-cell hopping along the $y$ axis, as the inter-edge distance is fixed. 
This limitation is overcome by manipulating the number of bonds (single as in B1 and B2 or double as in A1 and A2) that can be formed across the boundary of the unit cell. 

Together, the arrangements of zigzag egdes within the graphitic network lead to an effective tight-binding model characterized by an effective magnetic flux of $\pi$ in each $xy$ plaquette.
All four unit cells preserve mirror symmetry along $x$ ($M_x$), while $M_y$ is broken due to $J_x \neq J_x'$ (see inset in Fig.~\ref{fig:schematic}).
We treat this asymmetry as a perturbation and show in the Sec. III of Supplemental Material that it does not affect the topological phase. 
In the following, we neglect the distinction between $J_x$ and $J_x'$. 
The four unit cells can be classified (Table~\ref{tab:decision}) according to the relative strength of intra-cell and inter-cell hopping, a feature in analogy to the SSH model \cite{heeger1988solitons, asboth2016short, benalcazar_electric_2017, groning2018engineering}.

\begin{table}[h]
\centering
\begin{tabular}{c|c|c}
\toprule
 &$J_{x1}<J_{x2}$ & $J_{x1}>J_{x2}$  \\
\midrule
$J_{y1}<J_{y2}$ & \textbf{A1} & A2 \\
$J_{y1}>J_{y2}$ & B1          & B2 \\
\bottomrule
\end{tabular}
\caption{Comparison of effective hopping parameters in the four classes of graphitic structures A1, A2, B1, and B2. Structures are given in Fig.~\ref{fig:schematic}.}.
\label{tab:decision}
\end{table}

Density functional theory (DFT) is employed to calculate band structures of these structures (Fig.~\ref{fig:periodic}). 
Near the Fermi level, the band structures of the 2D graphitic lattice in Fig.~\ref{fig:schematic} resemble that of the simpler, effective tight-binding model (Fig. S1 in Supplemental Material) dominated by zigzag edges. 
Next, we explore the topology of the middle bands by calculating Wilson loop along $k_x$ and $k_y$ \cite{wang_higher-order_2019, benalcazar_quantized_2017, benalcazar_electric_2017, alexandradinata2014wilson}, respectively, for the two occupied bands (shaded blue) below the Fermi level. 
The logarithmic eigenvalues of the Wilson loops ($e^{i2\pi v_{x/y}}$) are plotted against $k_{y/x}$ and the resulting Wannier bands of A1 and B1 are shown in Fig. \ref{fig:periodic}, exhibiting gaps as a result of the negative hopping.  
In Fig. \ref{fig:periodic}c and \ref{fig:periodic}e, the presence of $M_x$ ensures that the two Wannier bands plotted against $k_y$ come in pairs ($v_x, -v_x$). 
The two Wannier bands $v_y$ would have also come in pairs regardless of the breaking of $M_y$ (Fig. S1), however, the two energy bands (shaded blue) below the Fermi level are inevitably coupled to low-lying bands, leading to distortion of Wannier bands depending on the strength of coupling.
Unit cell B1 has a larger gap between the blue-shaded bands and the bands below, thus the two bands are less coupled to low-lying bands than in A1 (Fig. S10). 
As a result, the $v_y$ bands of B1 remain nearly paired, whereas those of A1 exhibit noticeable deviations. 
This coupling does not affect the topological features, as evidenced by the existence of corner states in the numerical studies presented later (see also Sec. III of Supplemental Material).

\begin{figure}
\includegraphics[width=\columnwidth]{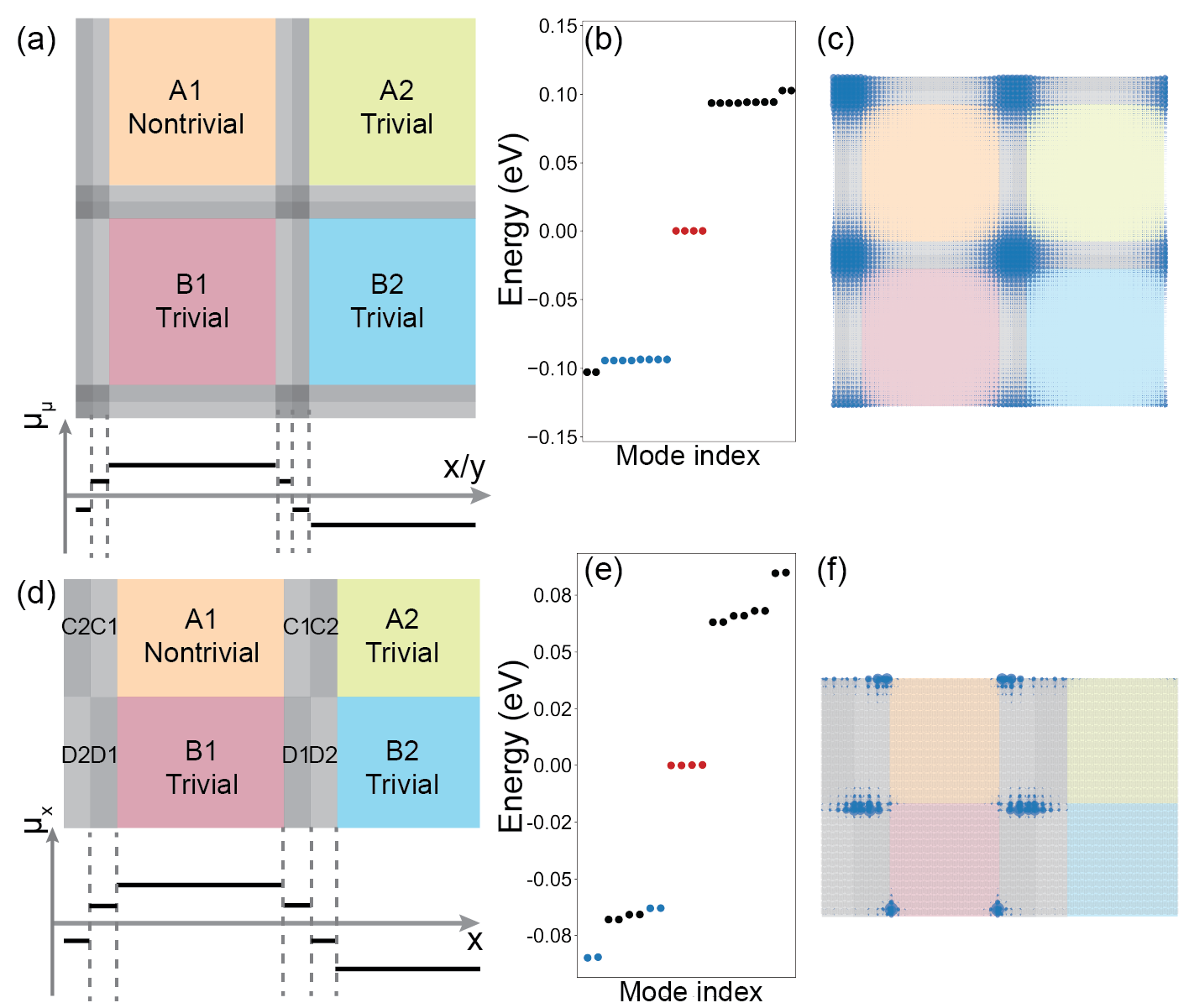}
\caption{\label{fig:corner_thick} (a) The analytical results are verified by numerically solving artificial tight-binding models constructed from unit cells with four basis sites. Domain walls with intermediate values of $\mu_\mu$ are inserted between adjacent regions along both $x$ and $y$ directions. (b) The spectrum of the artificial Hamiltonian exhibits four in-gap corner states (red) and sixteen additional localized states (blue, eight shown), in agreement with the analytical predictions. (c) Charge distribution of the sixteen massive states. (d) For graphitic structures, only $x$-directional domain walls can be adjusted, owing to structural constraints. (e) The resulting spectrum of graphitic structures displays four in-gap states (red) and four additional localized states (blue), whose charge distribution are given in (f). These results establish the existence of electronic codimension-2 states in a 2D material platform.
}
\end{figure}

\section{Edge Hamiltonian}
For second-order topological insulators, the bulk-boundary correspondence is encoded in the Wannier bands.
To reveal this correspondence of the Wannier bands, we consider a semi-infinite ribbon of the graphitic structure with open boundary conditions along $y$ and periodic boundary conditions along $x$ (Fig. \ref{fig:thin}a). 
Due to the large system size, we use tight-binding modeling assuming nearest-neighbor hopping between $p_z$ orbitals of carbon atoms rather than DFT to obtain wavefunctions, and show in the Sec. I of Supplemental Material that this approach achieves sufficient accuracy.
We then calculate Wilson loop for the $2N_y$ bands below the Fermi level and plot their Wannier centers ($v_x$). 
The iteration is until $2N_y$ because we only consider the middle two occupied bands below Fermi level. 
The results for A1 (Fig.~\ref{fig:thin}b) clearly show two degenerate Wannier edge states (red arrows) with quantized $v_x=0.5$. 
The Wannier edge states give rise to a quantized tangential polarization of $|e|/2$ (Fig. \ref{fig:thin}c). Similarly, the ribbon of A2 hosts degenerate edge states with $v_x=0.0$. In contrast, ribbons of B1 and B2 do not have edge states (Fig. S4-S6).

The existence of Wannier edge states is a manifestation of the nontrivial topology of Wannier bands in A1 and A2. It reflects the closing and reopening of their $v_x$ Wannier gap (Fig. \ref{fig:periodic}c) as an edge is created.
This is consistent with the assumption in Table~\ref{tab:decision} that A1 and A2 have $J_{y2}>J_{y1}$, rendering their $v_x$ bands nontrivial, whereas B1 and B2 have $J_{y2}<J_{y1}$ and therefore trivial $v_x$ bands.
Similar $y$-ribbon structures are also constructed with open boundary conditions along $x$ (Fig. S7-S8). The results suggest nontrivial $v_y$ bands for A1 and B1.
The topological invariant of these Wannier bands is the symmetry protected Wannier-sector polarization ($p_y^{v_x}$ and $ p_x^{v_y}$), which are calculated by nested Wilson loop approach based on both DFT and tight-binding modeling (Table S1-S2). 
The numerically quantized values further validate the conclusion here, i.e., unit cell A1 is a nontrivial quadrupole insulator with both Wannier bands $v_x$ and $v_y$ being nontrivial, as $q_{xy}=2e p^{v_x}_y p^{v_y}_x$. The above findings are summarized in Table~\ref{tab:class}.

\begin{table}[h]
\centering
\begin{tabular}{c|c|c}
\toprule
\diagbox{$v_x$ Wannier bands}{$v_y$ Wannier bands} & Nontrivial & Trivial \\
\midrule
Nontrivial & \textbf{A1} & A2 \\
Trivial  & B1          & B2 \\
\bottomrule
\end{tabular}
\caption{Topological classification of graphitic structures.}
\label{tab:class}
\end{table}

\section{Massless and massive corner states}

To reveal the topological corner state, we assemble the four distinct classes (A1, A2, B1, B2) together, as indicated by different colors in Fig. \ref{fig:thin}d. Each region contains $16 \times 16$ graphitic unit cells.
In this configuration, the second-order topological insulator A1 is surrounded by trivial phases. 
The supercell is then subject to periodic boundary conditions and the Hamiltonian is based on the tight-binding model of $p_z$ orbitals of carbon atoms. 
The resulting energy spectrum (Fig. \ref{fig:thin}e) demonstrates four degenerate in-gap states that are localized at the intersection of different domains (Fig. \ref{fig:thin}f) and correspond to the characteristic massless corner states of SOTIs.

Fig. \ref{fig:thin}d represents a thin domain wall between different classes where the parameter $\mu_\mu=J_{\mu1}-J_{\mu2}$ changes abruptly from positive to negative. 
To investigate the effects of the domain wall smoothness, we first construct an artificial lattice (Fig.~\ref{fig:corner_thick}a) comprising four regions, each belonging to different topological classes and containing $25 \times 25$ unit cells with four basis sites per unit cell. 
The boundaries between regions are smoothed by additional gray regions with intermediate value of $\mu_\mu$. 
Solving this artificial tight-binding Hamiltonian leads to the spectrum in Fig. \ref{fig:corner_thick}b, which exhibits the expected four in-gap states (red circles) localized at corners. 
More importantly, we find sixteen additional states below the Fermi level (blue circles, eight shown) that are also localized at the corners (Fig.~\ref{fig:corner_thick}c). 
Within the four corners of our configuration, the spectrum of localized states corresponds to one massless and four massive states per corner.
Their fourfold degeneracy is consistent with the analytical prediction, although the energy degeneracy is slightly broken in the current simulation due to discrete nature of tight-binding modeling. 
These states can be classified by angular momentum $l$ and exhibit spatial profiles analogous to atomic $s$ and $p$ orbitals, features that arise from the solutions of Eq.~\ref{eq:radial eq} (see details in Table~S6 and Sec.~IV of Supplemental Material). 
The massive states here are the two-dimensional generalizations of the VP states found in 1D.

To demonstrate the existence of codimension-2 states in real material systems, we now insert additional unit cells of graphitic structures into the configuration in Fig.~\ref{fig:thin}d to create a smooth domain wall for graphitic structures. 
Due to structural constraints, this is only feasible along the $x$ direction (Fig.~\ref{fig:corner_thick}d), leaving the domain wall along $y$ abrupt.
Nevertheless, the spectrum reveals four massless in-gap states (red circles) and four massive corner-localized states (blue circles, with charge distribution shown in Fig.~\ref{fig:corner_thick}f), corresponding to one massless state and one massive state per corner. 
The reduction from four to one massive state per corner, compared to the fully smooth case, is a consequence of the domain wall being smoothed along only one direction.
The features of the massive state, including number of degeneracy and weight distribution are consistent with results from an artificial lattice when only a single smooth domain wall is allowed (see details in Sec. IV and V of Supplemental Material).
These results demonstrate that codimension-2 states can be realized in 2D material systems by tuning the domain wall smoothness, even when smoothing is applied along only one direction.

\section{\label{sec:sec4} Conclusions}
In this work, we realize the complete phase diagram in quadrupole insulators and design four topological classes of 2D graphitic structures by engineering the arrangement of zigzag edges. 
With access to the full phase diagram, we create domain walls that are not limited by physical boundaries and whose smoothness can be controlled. 
We show that localized 0D topological boundary states emerge at the corner junction of the four classes, a result that is expected from the higher-order bulk-boundary correspondence. 
In addition, when the domain wall becomes smooth, massive localized states appear with energy that depends on the characteristic features of the domain wall; such states can even have nonzero angular momentum. 
Lastly, we briefly comment on the experimental feasibility of the design. The 2D graphitic structure shown here can be regarded as graphene nanoribbons being fused together. 
Previous studies \cite{sarker2024porous} have shown that closely spaced nanoribbons readily fuse through reactions at their edges, consistent with the design principle demonstrated here.

\textit{Acknowledgments.} This work is supported by Office of Naval Research through Multidisciplinary University Research Initiative (No. N00014-21-1-2537). This research used resources of the National Energy Research Scientific Computing Center (NERSC), a DOE Office of Science User Facility supported by the Office of Science of the U.S. Department of Energy under Contract No. DE-AC02-05CH11231. This work used computational and storage services associated with the Hoffman2 Cluster which is operated by the UCLA Office of Advanced Research Computing’s Research Technology Group.

\bibliographystyle{apsrev4-2}
\bibliography{cleaned}

\end{document}


\title{Supplemental Material for Massive Bound States in Second-order Topological Graphitic Structures}

\author{Haiyue Huang}
\affiliation{Division of Physical Sciences, College of Letters and Science, University of California, Los Angeles, 90095, California, USA}

\author{Prineha Narang}
\thanks{Contact authors}
\affiliation{Division of Physical Sciences, College of Letters and Science, University of California, Los Angeles, 90095, California, USA}
\affiliation{Department of Electrical and Computer Engineering, University of California, Los Angeles, 90095, California, USA}

\author{Ioannis Petrides}
\thanks{Contact authors}
\affiliation{Division of Physical Sciences, College of Letters and Science, University of California, Los Angeles, 90095, California, USA}

\maketitle


\section{\label{sec:computation} Computational details}
Density functional theory (DFT) calculations were performed with Vienna Ab initio Simulation Package (VASP) \cite{kresse1993ab,kresse1996efficiency,kresse1996efficient}. Perdew-Burke-Ernzerhof (PBE) based on Projector-Augmented-Wave (PAW) method pseudopotential was used. 
The k-grid, energy curoff and energy convergence criterion were set to $11\times11\times1$, 480 eV, and 1E-7 eV. Plotting of band structures was aided by VASPKIT \cite{VASPKIT}. The computation was repeated with Quantum Espresso \cite{giannozzi2009quantum} and was followed by Wannierization utilizing the Wannier90 \cite{mostofi2008wannier90} with a grid of $5\times5\times1$, and a number of disentanglement of 300. 
Maximally localized Wannier functions were then used to construct Wannier bands through Wilson loop calculations, as has been discussed in \cite{benalcazar_quantized_2017, wang_higher-order_2019}.

\begin{figure}
\includegraphics[width=0.9\columnwidth]{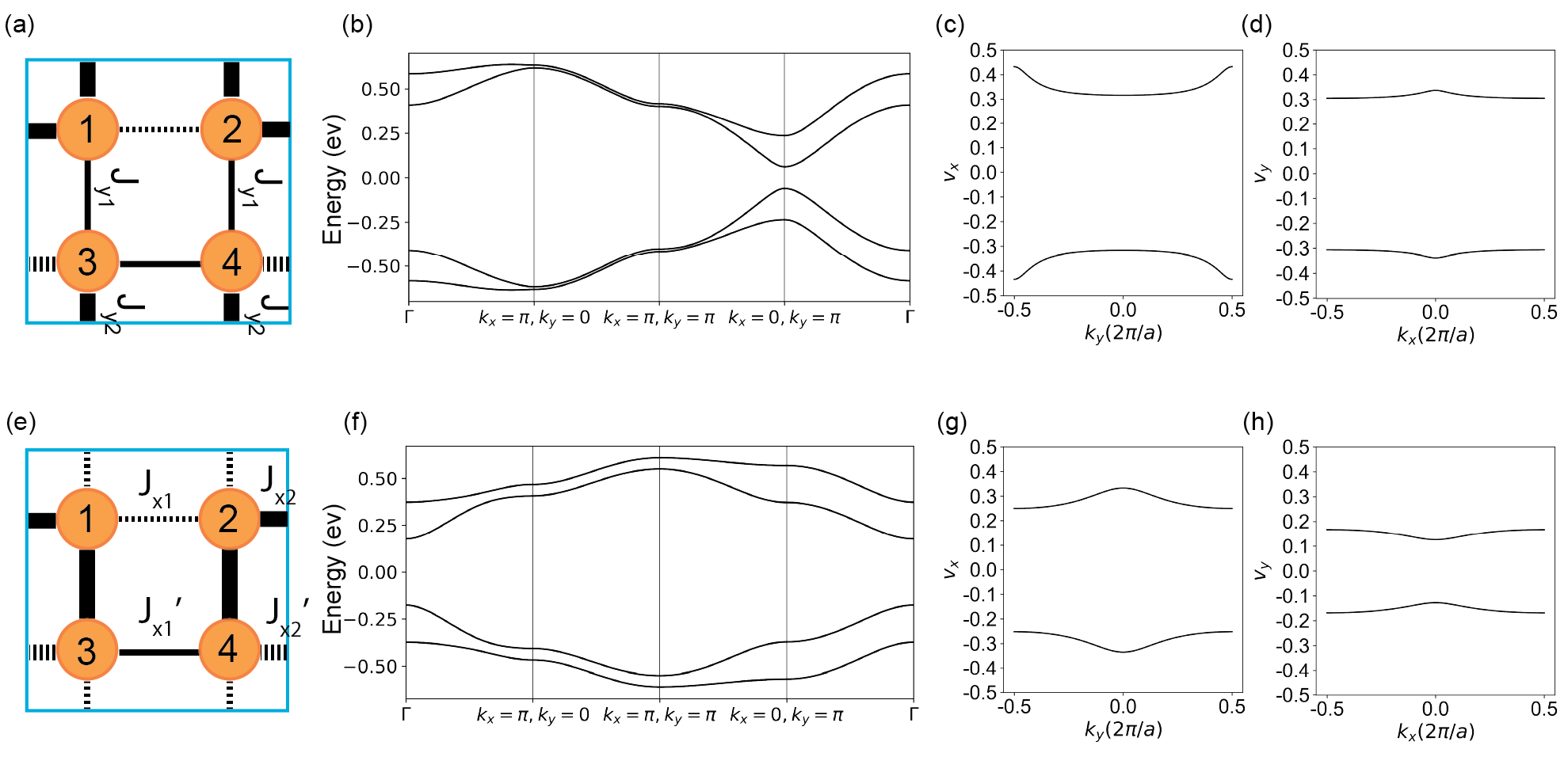}
\caption{\label{fig:simpleTB} (b) and (f) are band structures of the simplified, effective tight-binding models with four basis, which are schematically illustrated in (a) and (e), respectively. Solid and dashed lines indicate positive (negative) sign of the hopping strength. Thick (thin) lines indicate a larger (smaller) hopping strength. The results resemble those from DFT calculations (Fig. 2 in the main text). (c-d) and (g-h) are corresponding Wannier bands obtained by calculating Wilson loop on the two occupied bands.}
\end{figure}

Near the Fermi level, the Hamiltonian of graphitic structures can be simplified and effectively described by a simple tight-binding model with four basis sites (Fig. \ref{fig:simpleTB}) (hereafter referred to as the \textbf{effective Hamiltonian}). The hopping parameters respect the $M_x$ symmetry of the graphitic structures. The sign of hopping is alternatively positive (solid lines) or negative (dashed lines). Along the $x$ direction, the hopping strength between 1 and 2 ($J_{x}$), and between 3 and 4 ($J'_{x}$) can be different. Along the $y$ direction, the hopping strength between 1 and 3, and 2 and 4 are the same ($J_y$). By matching the gap width to that of DFT calculations, a set of hopping parameters can be determined (Table \ref{tab:TBparam}), which allows us to obtain the energy band structures (Fig. \ref{fig:simpleTB}b, \ref{fig:simpleTB}f).

The Hamiltonian can also be approximated by tight-binding calculations assuming nearest-neighbor hopping of \(p_{z}\) orbitals of carbon atoms (hereafter referred to as the \textbf{p-Hamiltonian}). This step was aided by the PythTB package \cite{yusufaly2013tight}. The nearest-neighbor hopping parameter was set to $-$2.7 eV. 

\begin{figure}
\includegraphics[width=0.9\columnwidth]{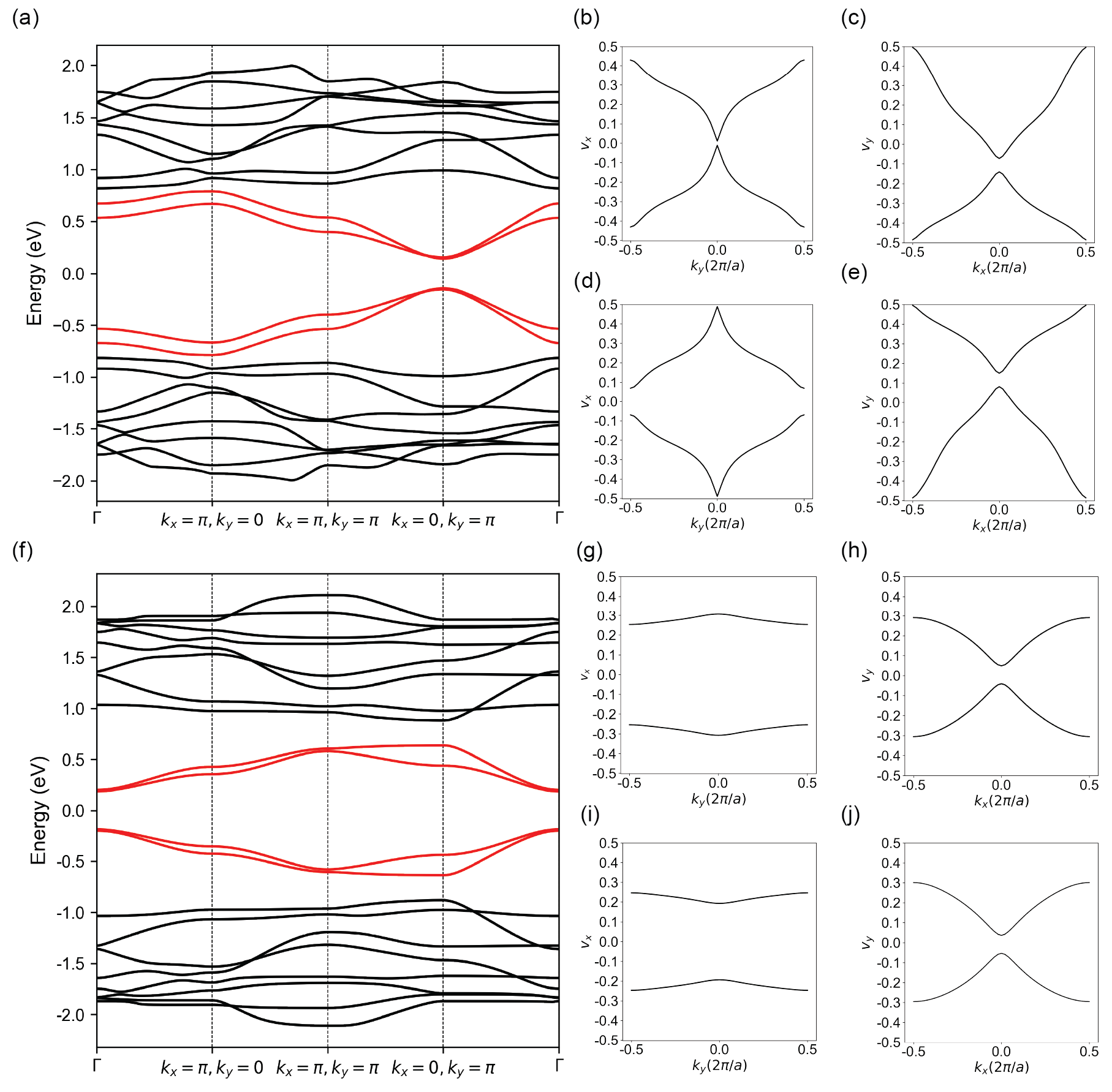}
\caption{\label{fig:TBband} (a) and (f) are band structures of unit cell A1/A2, and B1/B2, calculated with tight-binding models assuming nearest-neighbor hopping of \(p_{z}\) orbitals of carbon atoms (p-Hamiltonian). The middle bands (red) are consistent with those obtained with DFT methods. Wannier bands for A1 are obtained for the two occupied bands by calculating Wilson loop, either along (b) $k_x$ or (c) $k_y$ directions. Same Wannier bands are obtained for (d, e) A2, (g, h) B1, and (i, j) B2.}
\end{figure}

\begin{figure}
\includegraphics[width=0.9\columnwidth]{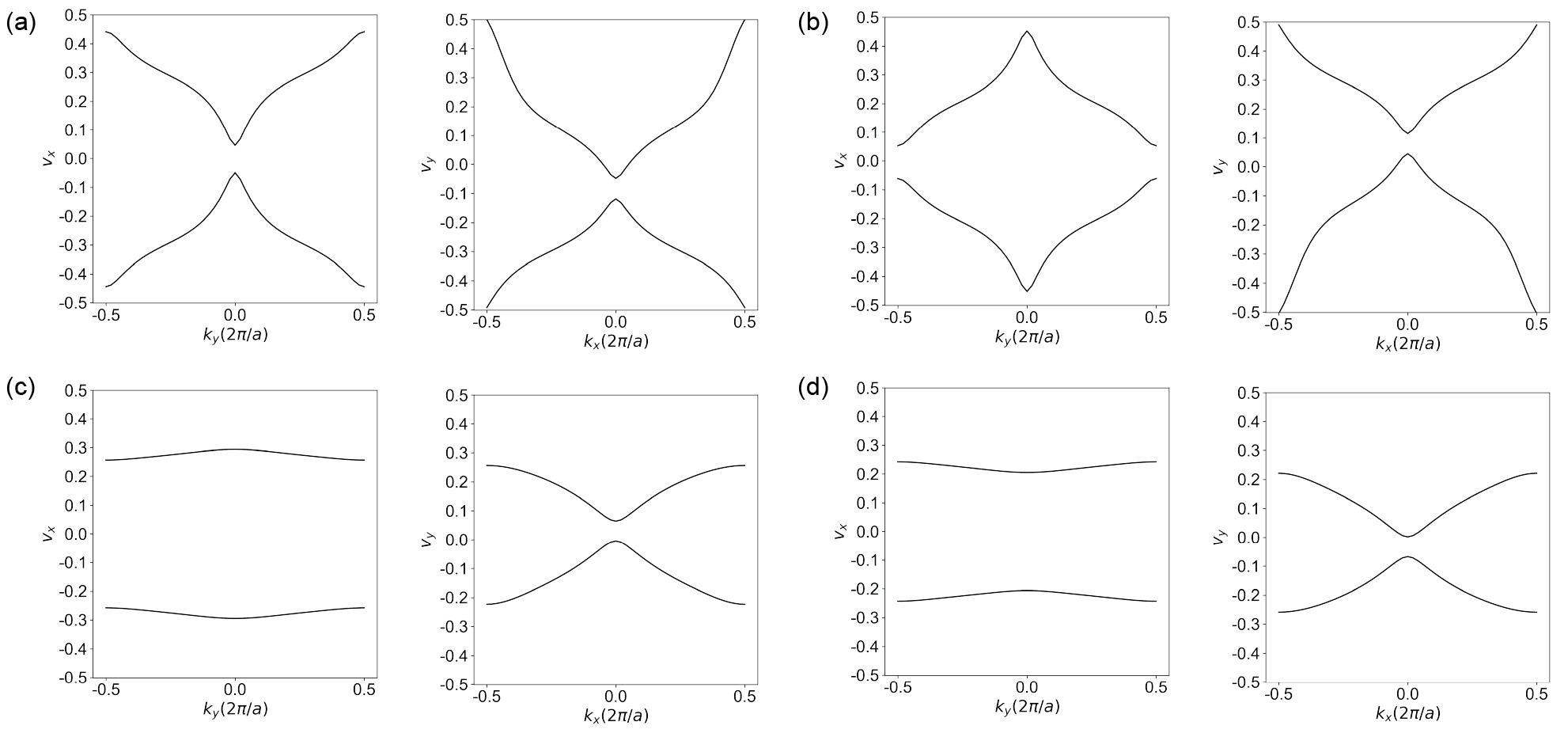}
\caption{\label{fig:Wband} Wannier bands obtained from maximally localized Wannier functions derived from DFT calculations for (a) A1, (b) A2, (c) B1, and (d) B2. They are consistent with results in Fig. \ref{fig:TBband}.}
\end{figure}

As demonstrated in Fig. \ref{fig:TBband}, the band structures calculated with the p-Hamiltonian are very similar to those obtained with DFT methods. By comparing Fig. \ref{fig:TBband} to Fig. \ref{fig:Wband}, it can be seen that the Wannier bands calculated with the two approaches are also very similar. This suggests that it is reasonable to use p-Hamiltonian for demonstrating edges states in ribbon structures and corner states when phase transition occurs.

\begin{figure}
\includegraphics[width=0.9\columnwidth]{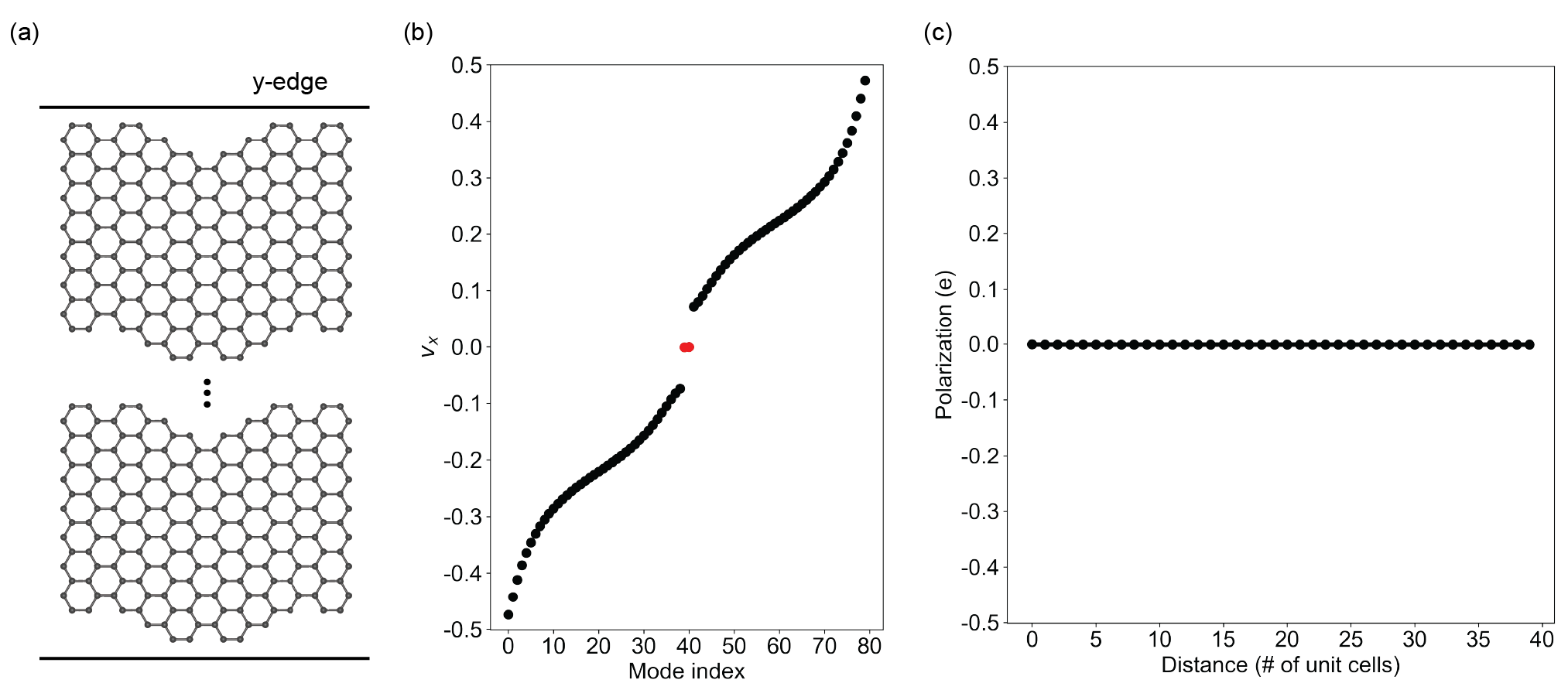}
\caption{\label{fig:ribbonW1} Same plot as Fig.~3a--3c but with unit cell A2. (a) Periodic $x$-ribbon structure made of unit cell A2. (b) Discrete values of $v_x$, showing two degenerate edge states with $v_x=0$. (c) This leads to zero tangential polarization.}
\end{figure}

\begin{figure}
\includegraphics[width=0.9\columnwidth]{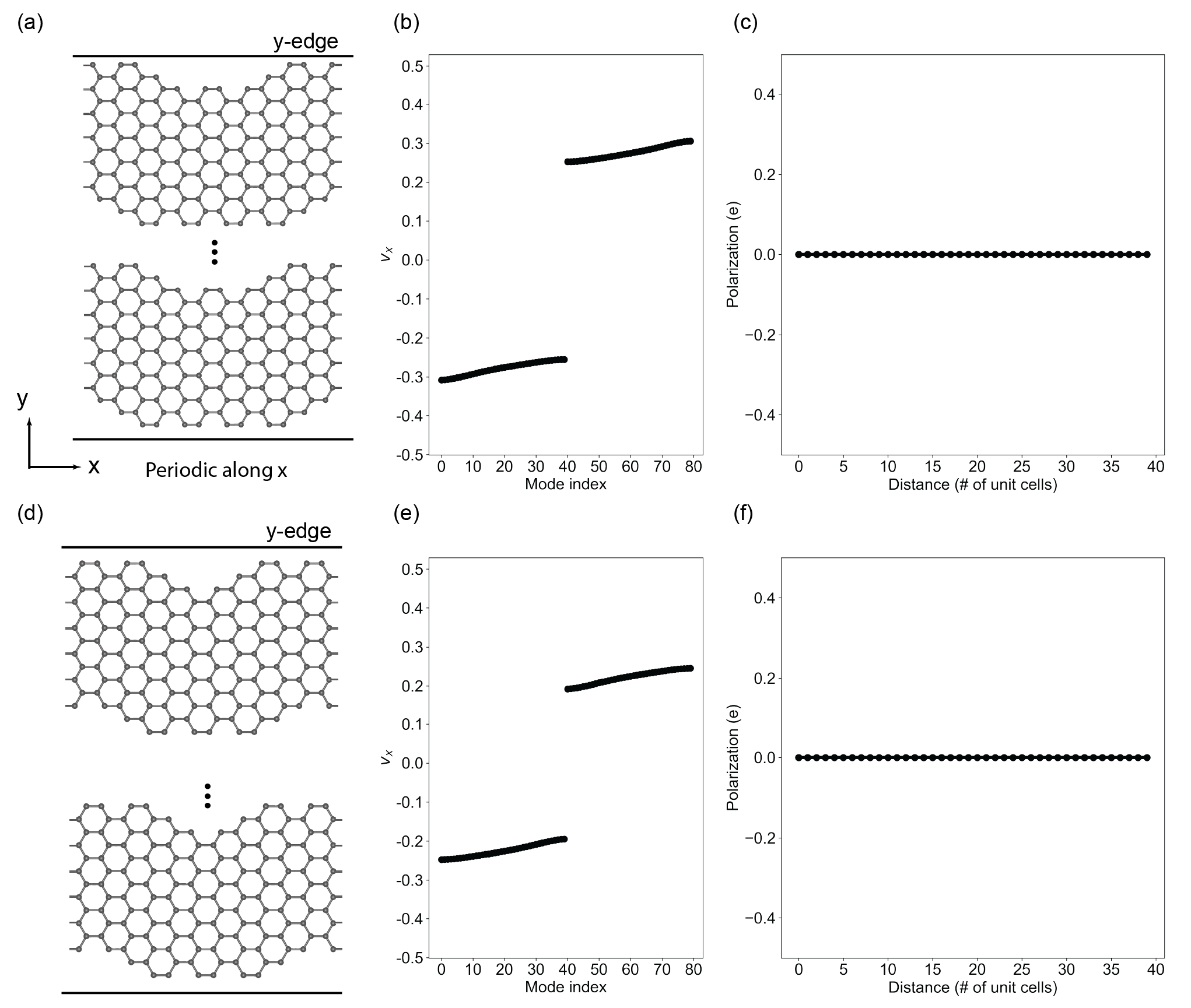}
\caption{\label{fig:ribbonW2} (a) and (d) are $x$-periodic ribbon structures made of unit cell B1 and B2, respectively. (b) and (e) are their corresponding discrete values of $v_x$, showing bulk values only. (c, f) As a result of the trivial bands, their tangential polarizations  are zero.}
\end{figure}

\begin{figure}
\includegraphics[width=0.6\columnwidth]{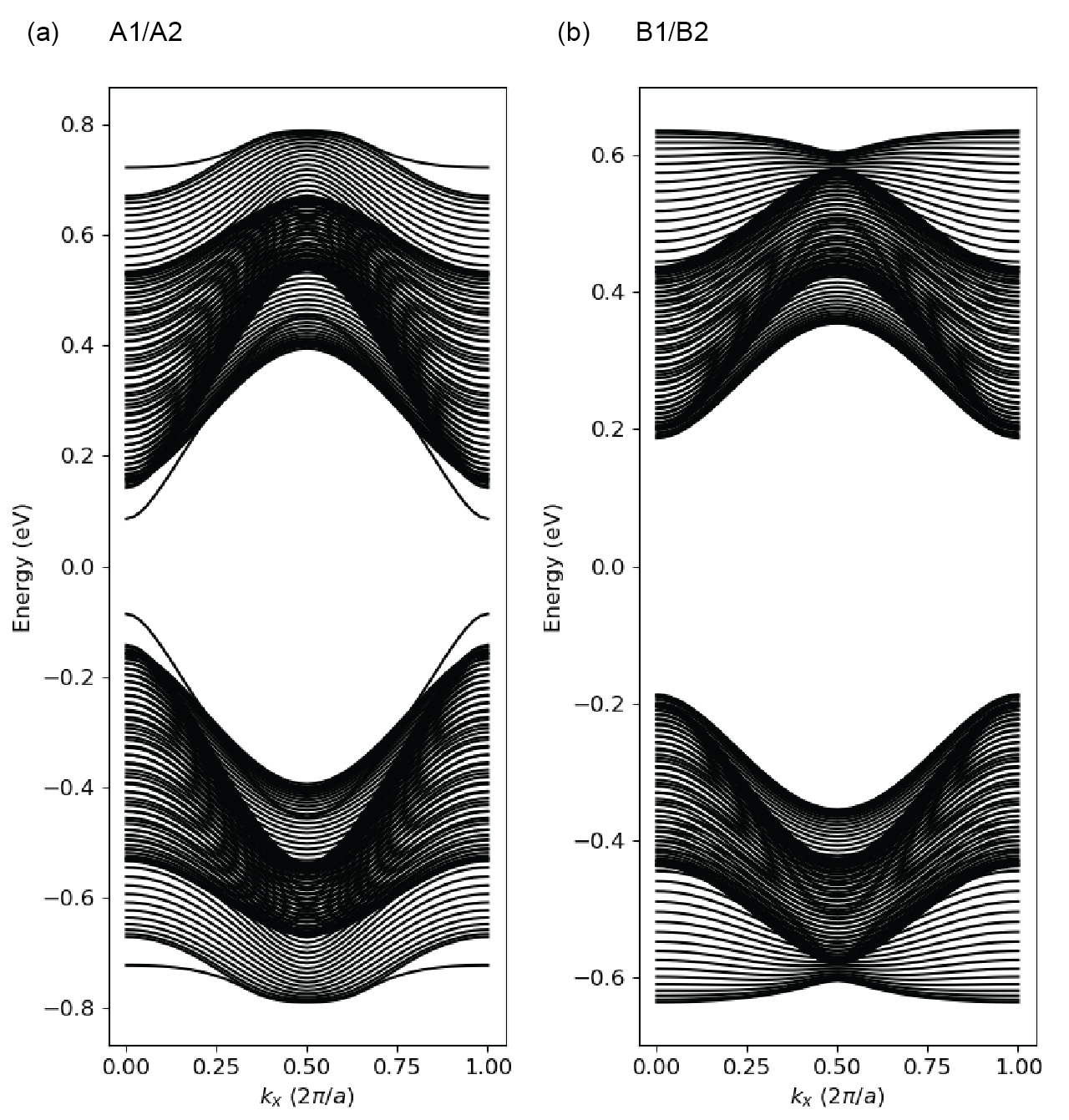}
\caption{\label{fig:ribbonBand} Band structures for the $x$-periodic ribbons made of (a) A1 and (b) B1 unit cells. Ribbons made of A2 (B2) unit cells share the same band structures with A1 (B1). Edge states exist in A1 and A2, but are absent in B1 and B2.}
\end{figure}

The $x$-periodic ribbon structure was constructed by creating a supercell with $N_y=40$ unit cells. It thus has an open boundary condition along $y$, and remains periodic along $x$. 
The Wilson loop calculation based on p-Hamiltonian yields Wannier values $v_x$. In obtaining the tangential polarization for ribbon structures, an onset potential of $\pm0.001$ eV was added to break the degeneracy.
For nontrivial Wannier bands, two Wannier edge states with quantized values of $v_x=0/0.5$ are expected. The case of $v_x=0.5$ has been demonstrated in structure A1 (Fig. 3) with tangential polarization of $|e|/2$. 
Here, the other case, $v_x=0.0$ is realized in A2 (Fig. \ref{fig:ribbonW1}). In contrast, the corresponding discrete Wannier spectrum $v_x$ for B1 and B2 (Fig. \ref{fig:ribbonW2}) does not yield degenerate edge states. 
Band structures of the ribbons are plotted in Fig. \ref{fig:ribbonBand} against $k_x$. For A1 and A2 (note they share the same ribbon band structures due to periodicity along $x$), their band structure yields edge states, which are separated from bulk bands. This again demonstrates the nontriviality of $v_x$ Wannier bands for A1 and A2.

\begin{figure}
\includegraphics[width=0.8\columnwidth]{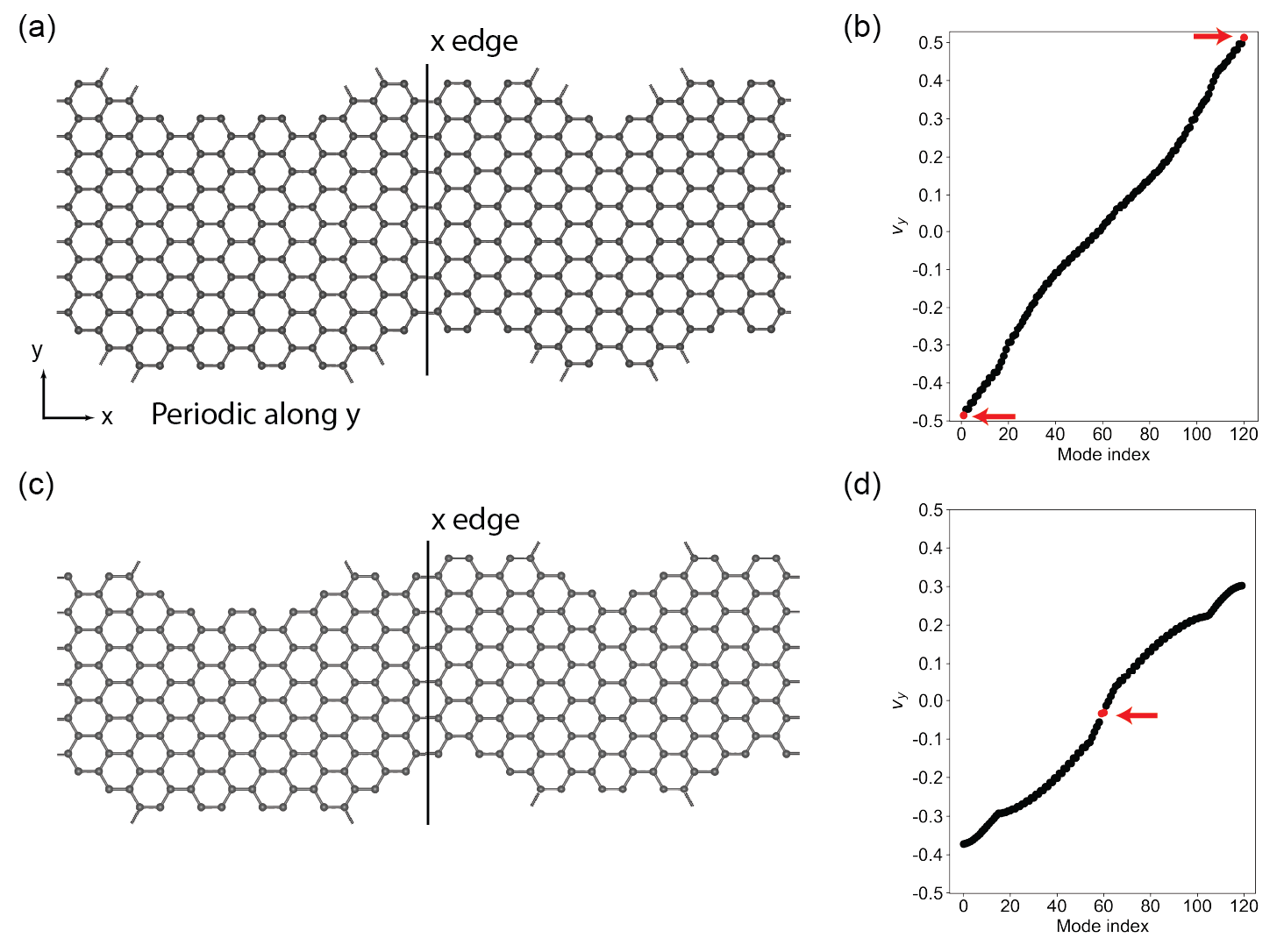}
\caption{\label{fig:ribbonWy} Periodic $y$-ribbon structures with open boundary conditions along $x$ can be constructed without introducing additional zigzag edges that would otherwise interfere with existing ones, by connecting (a) regions made by repeating unit cells A1 and A2, (c) B1 and B2, respectively. After computing Wilson loop, their $v_y$ are shown in (b) and (d), respectively, also showing pairs of degenerate edge states (red circle). The values of $v_y$ for these edge states are within the gap of Wannier bands (Fig. \ref{fig:TBband} and \ref{fig:Wband}).}
\end{figure}

\begin{figure}
\includegraphics[width=0.6\columnwidth]{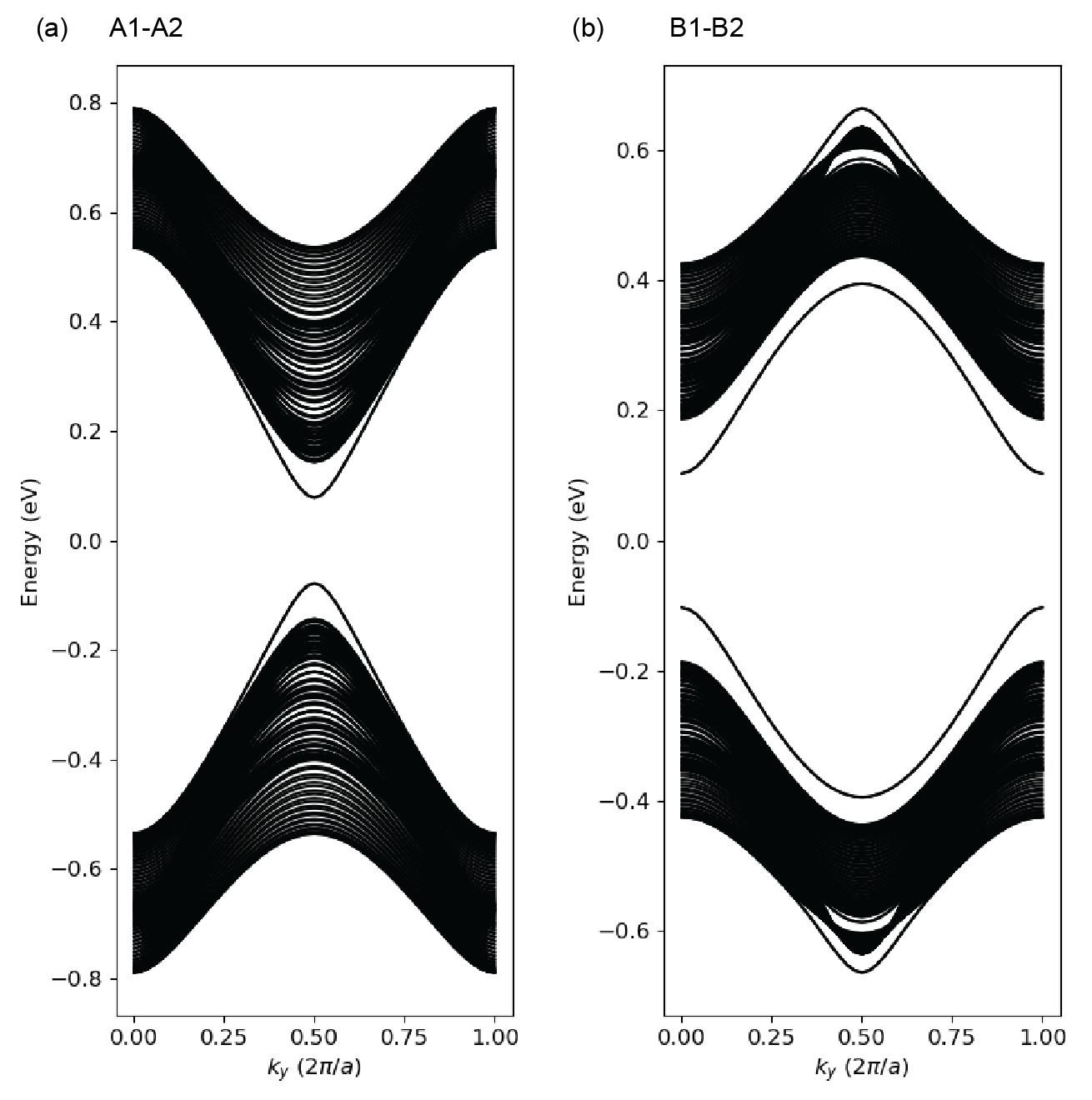}
\caption{\label{fig:ribbonBandy} Band structures for the $y$-periodic ribbons made by connecting (a) A1 and A2 together, and (b) B1 and B2 together. Both exhibit edges states separated from the bulk bands, as A1 (B1) and A2 (B2) belong to different topological classes. }
\end{figure}

Similar y-ribbon structures were also constructed with open boundary conditions along $x$. Instead of creating a physical boundary with vacuum, the interface was created by connecting two graphitic structures. Otherwise, additional zigzag edges would appear at the ends and interfere with the band structures.
The y-periodic ribbon structure was constructed by connecting $N_x=30$ unit cells of A1 and A2, or B1 and B2 together, making a total of 60 unit cells in the supercell. 
Fig. \ref{fig:ribbonWy}b and \ref{fig:ribbonWy}d show $v_y$ for the 2$N_x$ states in the composite ribbon of A1-A2, and B1-B2, respectively.
The existence of a pair of edge states (red dots) in Fig. \ref{fig:ribbonWy}b suggests a band gap closure and reopening,
and thus a transition of topological classes from the left (A1) to the right region (A2).
Similarly, the edges states in Fig. \ref{fig:ribbonWy}d signals B1 and B2 belong to different topological classes.
The band structures of ribbons show edge states (Fig. \ref{fig:ribbonBandy}a), which can come from either A1 or A2. Considering the simplified tight-binding picture where A1 has a stronger intracell cell hopping along $x$, here we assign the $v_y$ nontrivial class to A1. Likewise, the edge states in Fig. \ref{fig:ribbonBandy}b are assigned to come from B1. 
Altogether, the above analysis suggests A1 has nontrivial $v_x$ and $v_y$ Wannier bands, A2 has nontrivial $v_x$ and trivial $v_y$ bands. B1 has nontrivial $v_y$ and trivial $v_x$ Wannier bands. B2 has both bands trivial.

\begin{figure}
\includegraphics[width=0.6\columnwidth]{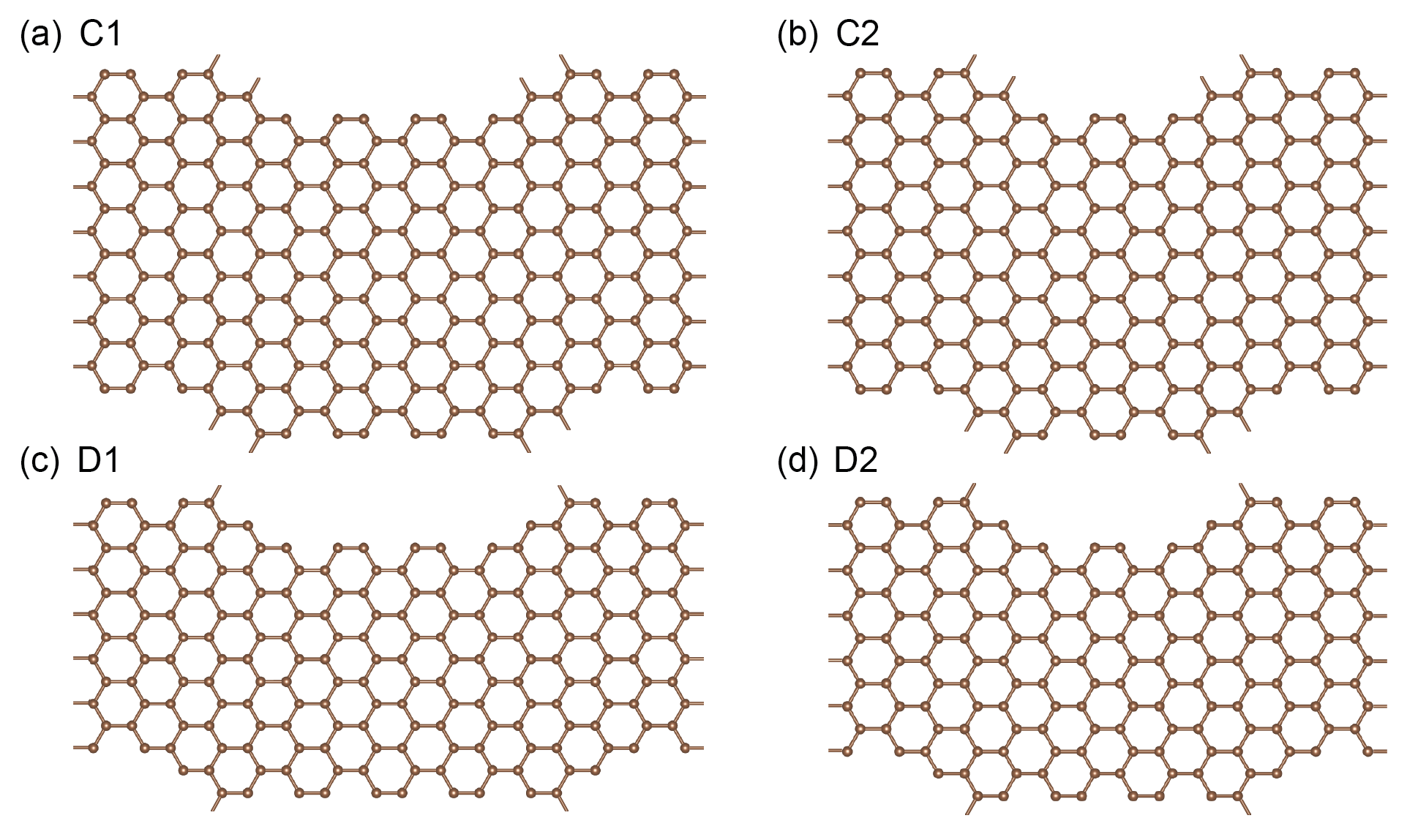}
\caption{\label{fig:additional} Additional unit cells used to create a smooth domain wall in Fig. 4. Carbon atoms on the edge are saturated with hydrogen atoms, which are not displayed.}
\end{figure}

To create a smooth domain wall, we inserted additional unit cells between regions made of A1 and A2, and between B1 and B2. These additional unit cells have intermediate values of $\mu_x$. The inserted gray regions in Fig. 4 are made of $2 \times 16$ unit cells of C1 and C2, D1 and D2, with structures drawn in Fig. \ref{fig:additional}.

\section{Bound state solutions}\label{appx:sol}
The components of the radial part $R_i(r)$ is found by the ansatz $R_i(r)=r^{|l|}e^{-\frac{\Lambda}{2\nu}r^2}f_i(r)$, leading to 
\begin{eqnarray}
    \begin{array}{rl}
       0=&  f_i''(r) + \left(\frac{2|l| +1}{r}-2\frac{\Lambda}{\nu} r\right)f_i'(r) \\&-\frac{1}{\nu^2}\left(\mu_0^2 - \epsilon^2+ 2\Lambda \nu(1+|l|+\Sigma_i)\right)f_i(r) 
    \end{array}
    \label{eq:difftrun}
\end{eqnarray}
where the subscript is hereafter dropped for simplicity.

To find an analytical solution, we use the Frobenius method and substitute $f(r)=\sum_{n=0}^\infty a_n r^{n+s}$ where $a_n$ are coeffiecients to be determined and $s$ is an unknown.
Substituting in Eq.~\ref{eq:difftrun} we obtain
\begin{eqnarray}
   \begin{array}{c}
      0=   a_n (n+s)(n+s+2|l|)+ \left(C-2\frac{\Lambda}{\nu}\left(n-2+s\right)\right)a_{n-2} 
   \end{array}
\end{eqnarray}
where $C=\frac{1}{\nu^2}\left(\epsilon^2-\mu_0^2 - 2\Lambda \nu(1+|l|+\Sigma_i)\right)$
The indicial equation is found by setting $n=0$, leading to ${s(s+2|l|)=0}$.
Since we are after normalizable solution we must choose $s=0$.
Hence the recurrence relation for the coefficients is
\begin{eqnarray}
        a_n =-\frac{C-2\frac{\Lambda}{\nu}\left(n-2\right) }{n\left(n+2|l|\right)} a_{n-2}
\end{eqnarray}
The above equation leaves $a_0$ unconstrained, while $a_1$ necessarily vanishes.
This leads to even power solutions for $f(r)$.
In order to obtain polynomial solutions the recurrence relation must end for some $n=2N+2$, leading to 
\begin{eqnarray}
   C-4\frac{\Lambda}{\nu}N =0
\end{eqnarray}
Substituting back the expression for $C$ and solving for the energy
\begin{eqnarray}
    \epsilon_{n,l} = \sqrt{\mu_0^2 +2\Lambda\nu (1+|l|+2n+\Sigma)}
\end{eqnarray}
In this case the wavefunctions are given in terms of the associated Laguerre polynomials 
\begin{eqnarray}
    f_{n,l}(r) = L^{(l)}_n (\frac{\Lambda}{\nu} r^2)
\end{eqnarray}

\section{Breaking mirror symmetry}
Here we consider the breaking of mirror symmetry by adding $h'$ to Hamiltonian 
\begin{align}
 \tilde{\hat{h}}=& \sum_{{\mathbf{m}}}\left[\hat{H}_{x}(\tilde{{m}} )+\hat{H}_{y}(\tilde{\mathbf{m}}) + \Delta\hat{H} _{xy} ({\mathbf{m}})\right]+\hat{h'}\,,
\label{HamProdK}
\end{align}

\begin{align}
\hat{h'}=& \Delta J^- c^\dagger_{{\mathbf{m_x}}+\mathbf{e}_x}c^{}_{{\mathbf{m_x}} } + \Delta J^+ c^\dagger_{{\mathbf{m_x}}-\mathbf{e}_x} c^{}_{{\mathbf{m_x}}  } \,,\label{eq:Creutz}
\end{align}

This breaks the $M_y$ symmetry. We seek wavefunctions of the form 
\begin{align}
 \lvert \tilde{n} \rangle \approx
\lvert n \rangle +
\lvert n' \rangle ,
\end{align}

The first-order corrections can be written as a sum of unperturbed states,

\begin{align}
 \lvert u_{n}' \rangle = \sum_{m} c_{mn} \lvert u_m \rangle ,
\end{align}

\begin{align}
    c_{mn}=\frac{\langle u_m \rvert \hat{h'} \lvert u_n \rangle}{E_n-E_m} \equiv \frac{h'_{mn}}{E_n-E_m} \quad \text{for } m \neq n.
\end{align}

And $c_{nn}=0$. We now calculate the first-order corrections to the Wilson loop of the two occupied energy bands. We follow the definition discussed before \cite{wang_higher-order_2019,benalcazar_quantized_2017} 
\begin{align}
[W]_{nm}= \langle u_n (k_0) | V \Pi(k_0)|u_m(k_0)\rangle
\end{align}

\begin{align}
\Pi=P(k_0+2\pi)P(k_0+\frac{2\pi(N-1)}{N})...P(k_0+\frac{2\pi}{N})
\end{align}
\begin{align}
    \tilde{P}(\mathbf{k})=\sum_{n=1}^{n_{occ}}|\tilde{u}_n(\mathbf{k}) \rangle \langle \tilde{u}_n(\mathbf{k}) | 
\end{align}
\begin{align}
[V]_{nm}=|u_n(k_0)\rangle \langle u_m(k_0+2\pi)|
\end{align}

In our case, we have two occupied bands and they are degenerate when the Hamiltonian is unperturbed. Under $h'$ the nonzero weights are $c_{13}$ and $c_{31}$. Therefore,
\begin{align}
|\tilde{u}_1 \rangle \langle \tilde{u}_1 | = |u_1 \rangle \langle u_1 | + c_{31} (|u_1 \rangle \langle u_3 |+|u_3 \rangle \langle u_1 |) + c_{31}c_{31}|u_3 \rangle \langle u_3|
\end{align}

If we consider $\Delta J$ in the same order of magnitude of $J$, it turns out that $c_{mn}$ is of order $10^{-3}$. We thus neglect the contribution from the change of projector $P$. Meanwhile, when the basepoint is set to the gamma point, $c_{13}=c_{31}=0$ and thus $|u_n(k_0=0)\rangle$ remain unchanged under the perturbation of $h'$. Therefore, we conclude that breaking mirror symmetry ($M_y$) by changing the hopping parameters along $x$ does not influence the quadrupole. This can be verified by the effective Hamiltonian (Fig.~\ref{fig:simpleTB}c-d, ~\ref{fig:simpleTB}g-h) where the Wannier bands remain paired and gapped, and the nested Wilson loop still has quantized eigenvalues, regardless of the distinction between $J_x$ and $J'_x$.

\begin{figure}
\includegraphics[width=0.7\columnwidth]{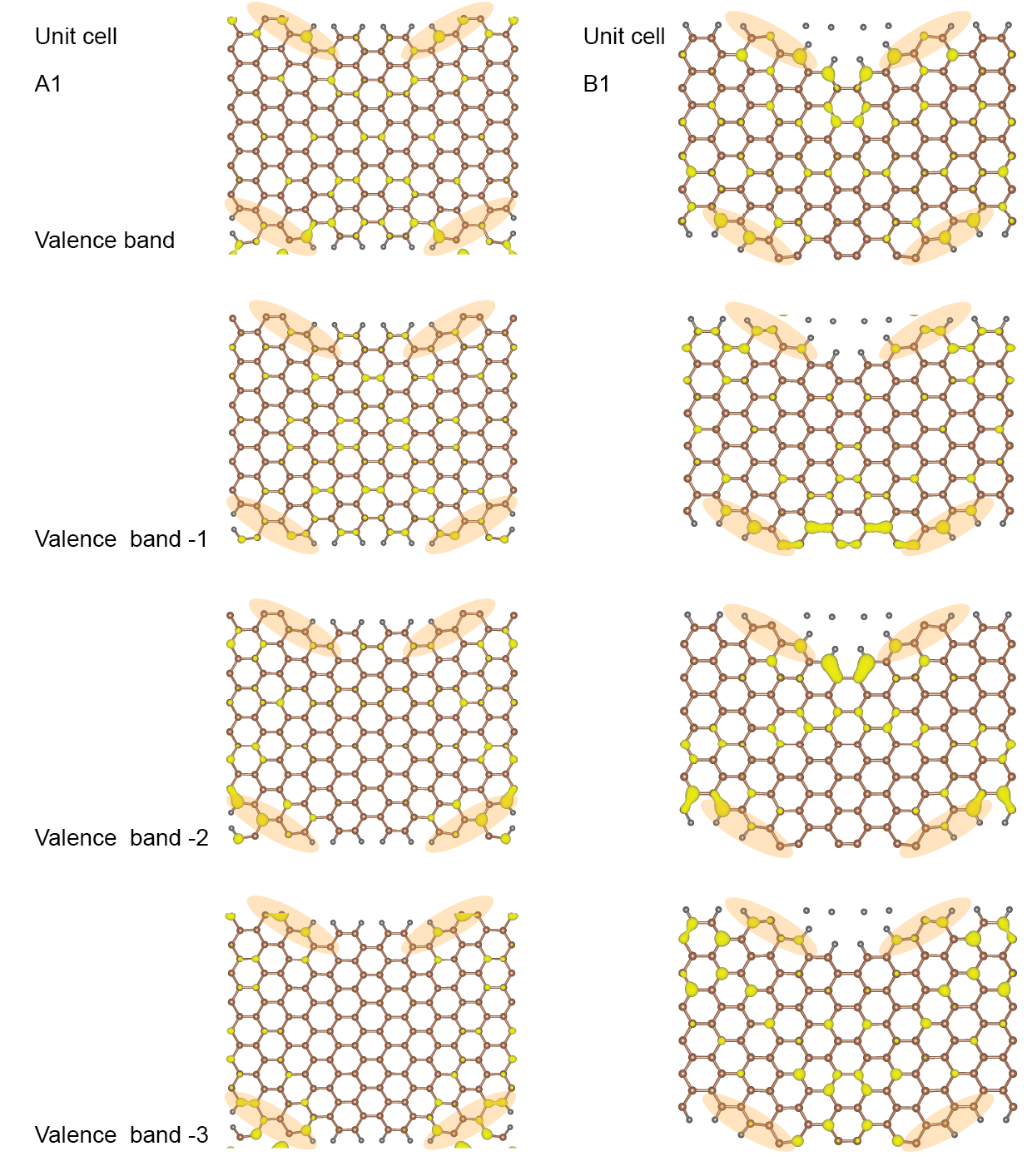}
\caption{\label{fig:coupledBands} The left and right columns are wavefunctions of A1 and B1 at gamma point for the four bands below the Fermi level, respectively. In A1, the lower two bands (valence bands -2 and -3) exhibit weight on zigzag edges, indicating bands coupling with the upper two bands, whereas such coupling is much weaker in B1.
}
\end{figure}

However, as discussed in the main text, the $v_y$ Wannier bands for structure A1 and A2 are not paired. This discrepancy arises because the analysis assumes a total of four bands with two occupied. In graphitic structures, however, the two bands below the Fermi level inevitably hybridize with low-lying bands, causing $v_y$ to deviate. The coupling with low-lying bands can be visualized by plotting the distribution of wavefunctions (Fig. \ref{fig:coupledBands}). For unit cell A1/A2, the band ``valence$-2$" and ``valence$-3$'' has obvious weight on the zigzag edges (orange shades), whereas such phenomenon is less obvious in B1/B2. As a result, we see the two $v_y$ Wannier bands are more symmetric for B1/B2 compared to A1/A2 (Fig. 2).

Nested Wilson loop was calculated as described in \cite{benalcazar_quantized_2017, benalcazar_electric_2017}. Polarizations over the Wannier sector with a base point of $k_x=-\pi/a$, $k_y=-\pi/a$ (where $a$ is the lattice constant) are summarized in Table \ref{tab:pxpy} and \ref{tab:pxpy2}. 
The results, together with the fact that in-gap corner states are observed, lead to the conclusion that coupling to low-lying bands does not influence the topological property of graphitic structures.

\section{Artificial lattice demonstrating massive states}
\begin{figure}
\includegraphics[width=0.9\columnwidth]{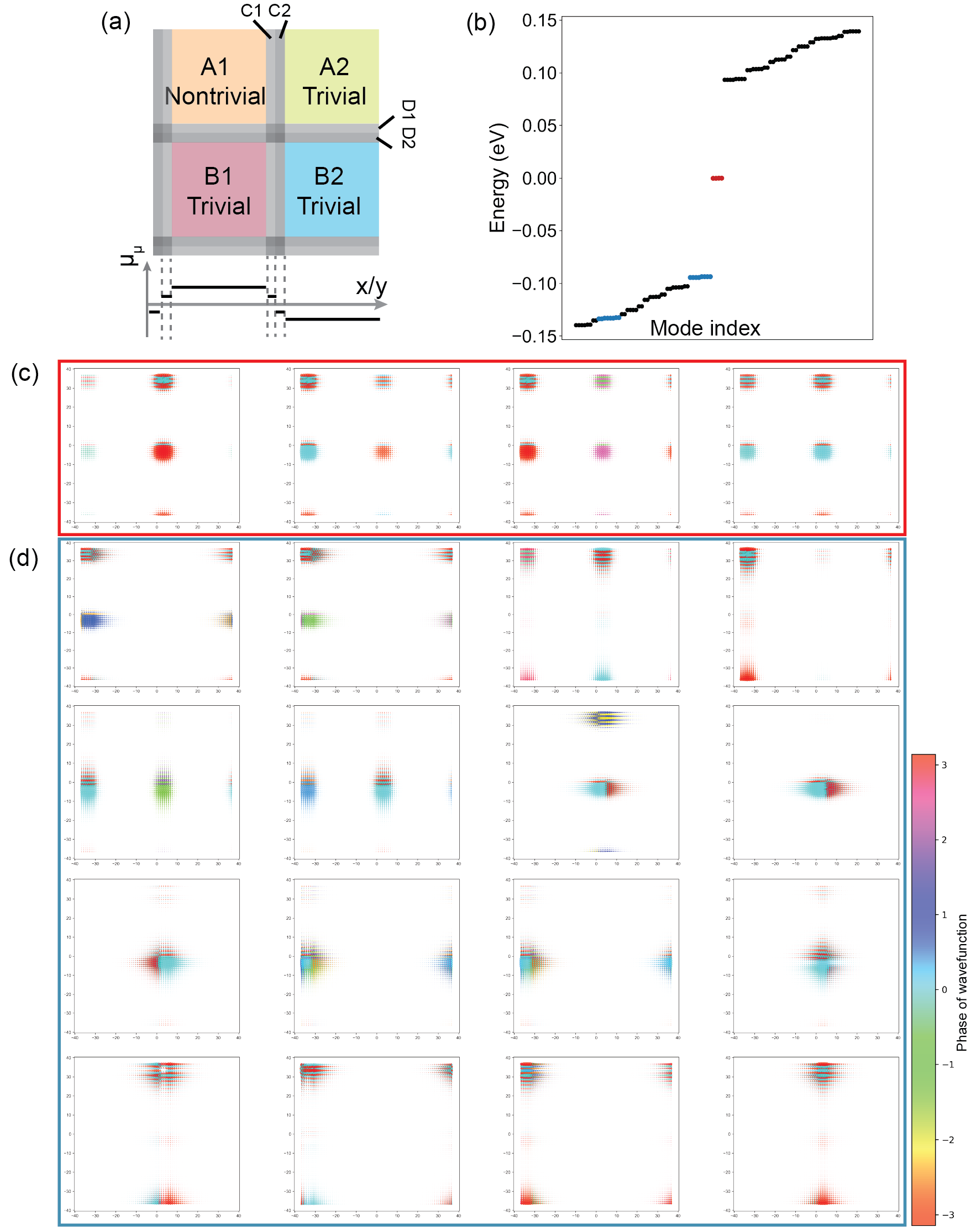}
\caption{\label{fig:TBartificial} (a) Schematic of the artificial tight-binding model in real space, where each unit cell contains four basis sites coupled by artificially chosen hopping parameters (see Table \ref{tab:TBparam3}) to examine analytical results (same plot as Fig. 4a in the main text). (b) Spectrum of the Hamiltonian, showing four massless states (red) and all sixteen massive states (blue). (c) and (d) are wavefunctions of the four degenerate massless states (red box) and sixteen massive states (blue box). These correspond to one massless state and four massive states per corner. The color and radius of each circle in the plot represent phase and intensity of the wavefunction, respectively.
}
\end{figure}

To build an artificial lattice with smooth domain walls along both directions, we constructed four regions, each having $25 \times 25$ unit cells with four basis sites (hereafter referred to as the \textbf{artificial Hamiltonian}). The set of hopping parameters used is listed in Table \ref{tab:TBparam3}. 
As shown in Fig. \ref{fig:TBartificial}b, the smoothness of the domain wall does not affect the energies of massless states. A total of sixteen massive states are identified, all lying below the Fermi level (blue circles in Fig. \ref{fig:TBartificial}b) and localized at the corners (Fig. \ref{fig:TBartificial}d). 

\begin{figure}
\includegraphics[width=0.95\columnwidth]{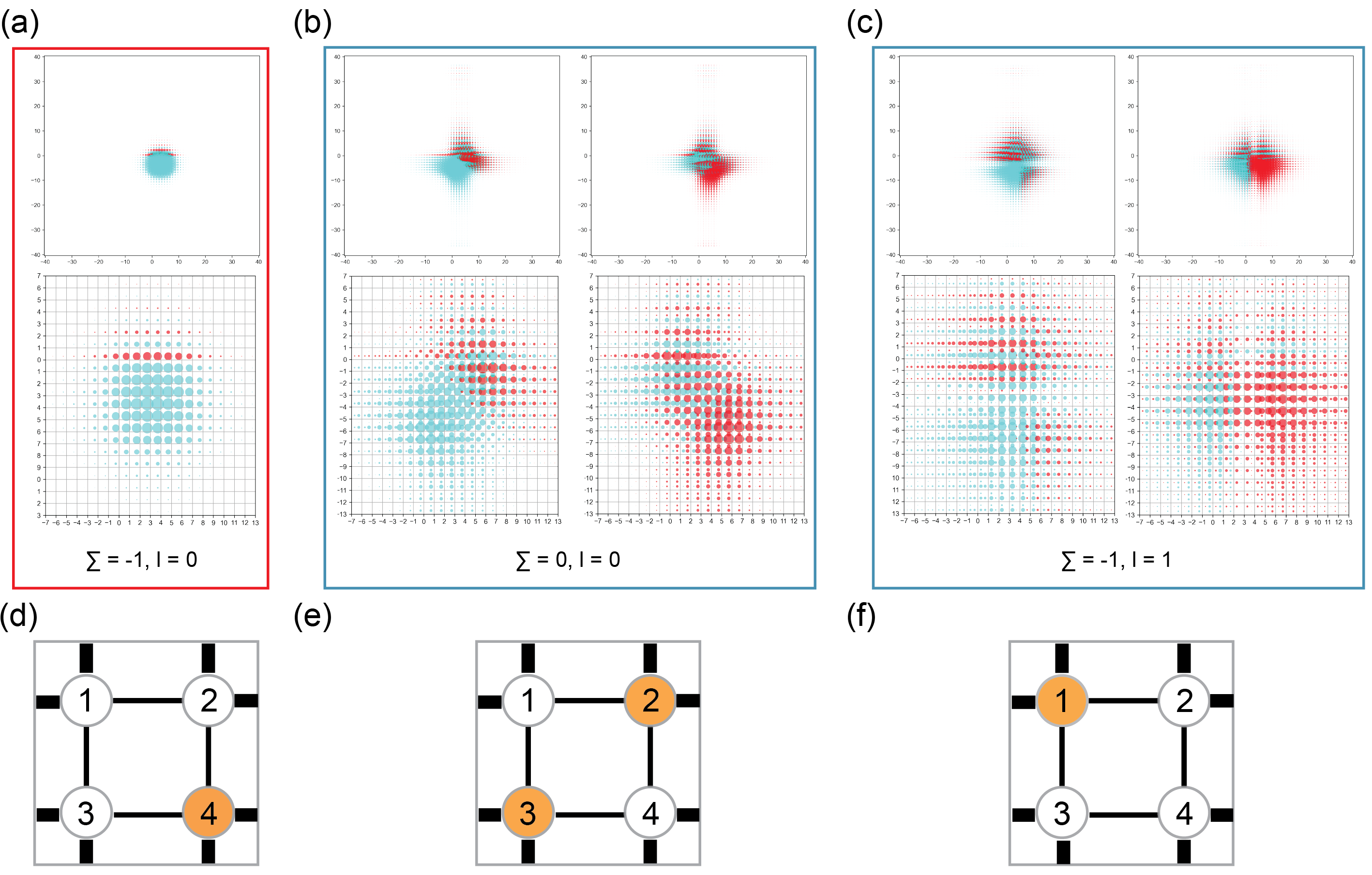}
\caption{\label{fig:spOrbital} (a)--(c) Replot of wavefunctions in Fig.~\ref{fig:TBartificial}, but with small onsite term added to lift the degeneracy. The lower row shows zoomed-in views, where each square represents one unit cell consisting four basis sites (No. 1--4). (d) Within each unit cell, we can investigate the weight distribution of wavefunctions to determine their $\Sigma_{ii}$.
For the massless states localized at the central corner, the wavefunction is weighted on basis site No. 4, which is consistent with the only possible value of $\Sigma_{ii}=-1$ as listed in Table.~\ref{tab:degeneracy}.
For the massive states, the wavefunction distributions exhibit two distinct patterns, corresponding to (e) $\Sigma_{ii}=0$ and (f) $\Sigma_{ii}=-1$, which are associated with angular momenta $l=0$ and $l=1$ as listed in Table~\ref{tab:degeneracy}, consistent with the (b) $s$-orbital-like and (c) $p$-orbital-like wavefunctions, respectively. These numerical results are in good agreement with the analytical predictions.
}
\end{figure}

According to Table \ref{tab:degeneracy}, massless state has the lowest energy (zero energy).
The states with second lowest energy are four-fold degenerate. They can be further classified according to their angular momentum $l=0$ or 1 and have a tendency on the weight distribution over basis sites, as governed by $\Sigma_{ii}$ (Eq. 9). These features are all observed in the numerical studies. The results show one massless state and four massive states per corner, though the degeneracy of massive states is slightly broken due to the discrete nature of tight-binding modeling.
As demonstrated in Fig.~\ref{fig:spOrbital}b-c, the spatial distributions of massive states are analogous to atomic $s$ and $p$ orbitals. We can then assign their angular momentum to be $l=0$ and $l=1$, respectively. According to Table~\ref{tab:degeneracy}, this then means that their $\Sigma_{ii}$ should be 0 and $-1$, respectively, which is consistent with the observed weight distribution (Fig. \ref{fig:spOrbital}d-f). Therefore, the existence of massive states in 2D is confirmed in numerical studies based on artificial lattices.

\section{Connection between the artificial lattice and the graphitic structure}

We now establish the connection among the artificial Hamiltonian, the effective Hamiltonian, and the p-Hamiltonian, where the latter provides a more accurate description of the graphitic structures.

\begin{figure}
\includegraphics[width=0.95\columnwidth]{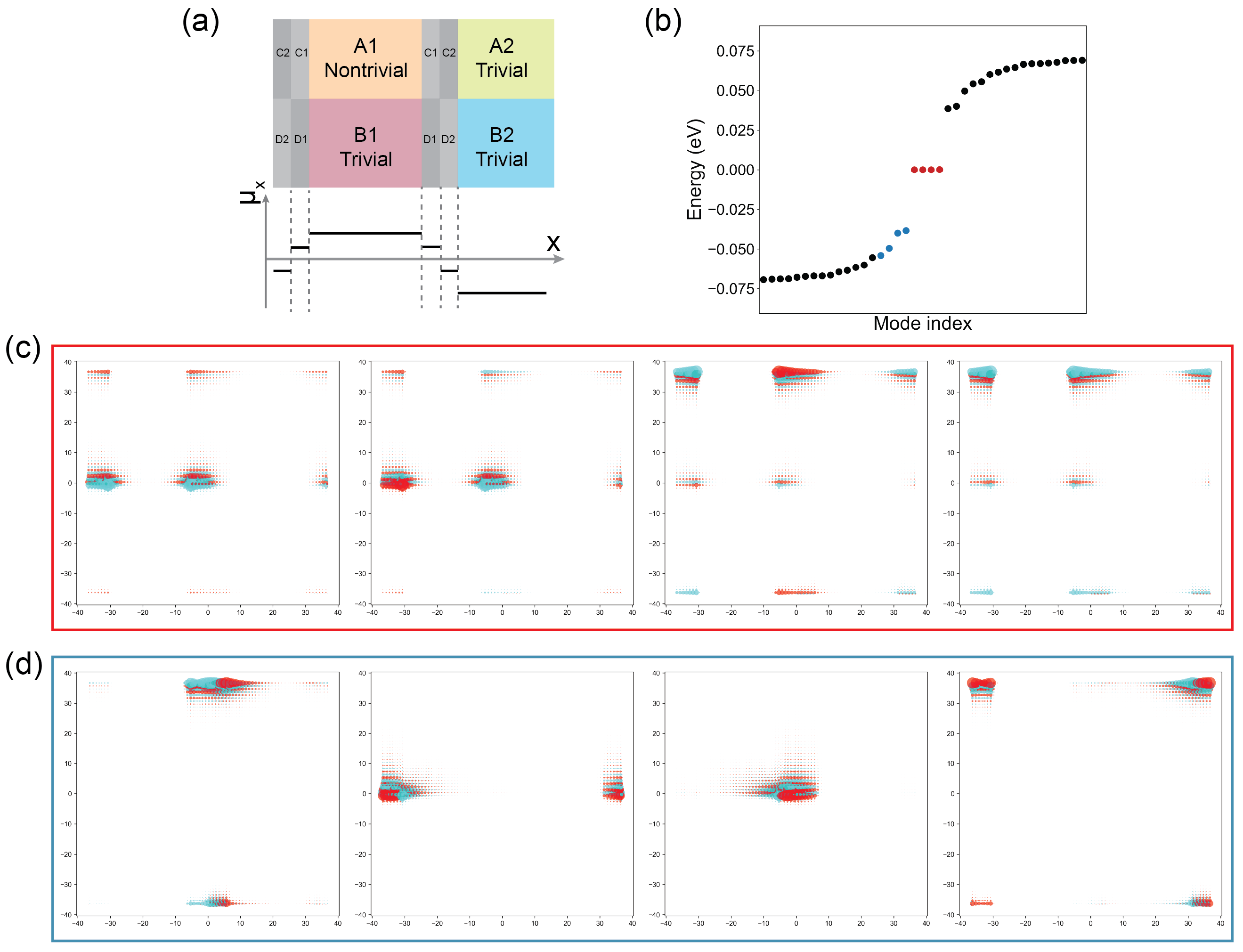}
\caption{\label{fig:TBGnr} (a) Schematic of the effective tight-binding model constructed from unit cells containing four basis sites, coupled by effective hopping parameters of zigzag edges in graphitic structures (Table \ref{tab:TBparam}). The layout follows the same format as Fig. 4d, but each unit cell contains only four basis sites. (b) The spectrum demonstrates four gapless states (red) and four massive states (blue), which is consistent with the results in a more complex graphitic system.
(c) and (d) are wavefunctions of the four degenerate massless states (red box) and massive states (blue box). All states are localized at the corners. The color and radius of each circle in the plot represent phase and intensity of the wavefunction, respectively.
}
\end{figure}

\begin{figure}
\includegraphics[width=0.95\columnwidth]{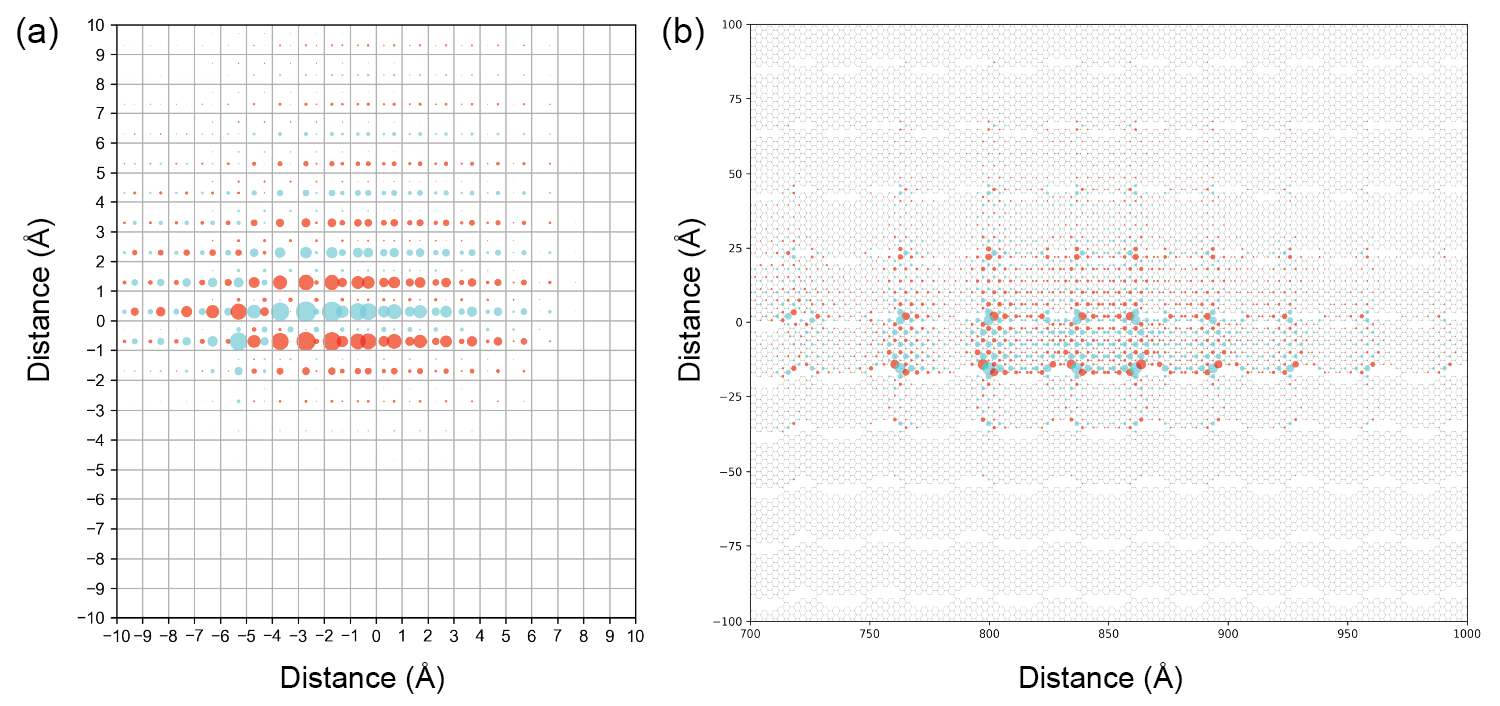}
\caption{\label{fig:compareCorner} (a) Zoomed-in view of the central corner state derived from the effective Hamiltonian (Fig.~\ref{fig:TBGnr}). (b) Central corner state of graphitic structures (showing only carbon atoms), derived from the p-Hamiltonian. It has a similar distribution of wavefunction as (a).  }
\end{figure}

To this end, we created similar four-region structures using the effective hopping parameters (Table \ref{tab:TBparam}) and add unit cells with intermediate $\mu_\mu$ values (Table \ref{tab:TBparam2}) to form a smooth domain wall along the $x$ direction (Fig. \ref{fig:TBGnr}b). This mimics the situation of domain walls in graphitic structures. Meanwhile, this also corresponds to a special case of the artificial Hamiltonian in which only one direction of domain wall is allowed, thus connecting the results of the artificial Hamiltonian and the p-Hamiltonian. The results in Fig. \ref{fig:TBGnr} show four massless states and four massive states. The number of states is consistent with those derived from the p-Hamiltonian in Fig. 4d-4e. Moreover, if we zoom in to check the weight on lattice sites, we find that the tendency of wavefunction distribution on lattice sites in the massive state from the effective Hamiltonian matches that from the p-Hamiltonian (Fig \ref{fig:compareCorner}). 
These results confirm that the additional localized states identified in graphitic structures are indeed massive states, exhibiting the features predicted in analytical studies.

\bibliographystyle{apsrev4-2}
\bibliography{cleaned}

\onecolumngrid

\begin{table}
\centering
\begin{tabular}{c c c c c c c}
\toprule
      & $p^{v_x}_y$ & $p^{v_y}_x$ \\ 
\midrule      
A1    & 0.495 & 0.500 \\ 
A2    & 0.495 & 0.000 \\ 
B1    & 0.007 & 0.500 \\ 
B2    & 0.007 & 0.000 \\ 
\bottomrule
\end{tabular}
\caption{Wannier-sector polarizations based on tight-binding modeling of $p_z$ orbitals on carbon atoms (\textit{i.e.} p-Hamiltonian).}
\label{tab:pxpy}
\end{table}

\begin{table}
\centering
\begin{tabular}{c c c c c c c}
\toprule
      & $p^{v_x}_y$ & $p^{v_y}_x$ \\ 
\midrule
A1    & 0.498 & 0.486 \\ 
A2    & 0.498 & 0.014 \\ 
B1    & 0.015 & 0.497 \\
B2    & 0.015 & 0.003 \\
\bottomrule
\end{tabular}
\caption{Wannier-sector polarizations based on DFT.}
\label{tab:pxpy2}
\end{table}

\begin{table}
\centering
\begin{tabular}{c c c c c c c}
\toprule
 & $J_{x1}$ & $J_{x2}$ & $J_{x1}'$ & $J_{x2}'$ & $J_{y1}$ & $J_{y2}$ \\
\midrule
A1 & 0.0871 & $-0.3082$ & $-$0.1821 & 0.2281 & $-$0.2067 & $-$0.2710 \\
A2 & 0.2281 & $-0.1821$ & $-$0.3082 & 0.0871 & $-$0.2067 & $-$0.2710 \\
B1 & 0.0320 & $-$0.3339 & $-$0.1607 & 0.2657 & $-$0.3044 & 0.1196 \\
B2 & 0.2657 & $-$0.1607 & $-$0.3339 & 0.0320 & $-$0.3044 & 0.1196 \\
\bottomrule
\end{tabular}
\caption{Tight-binding parameters for the effective Hamiltonian (Fig. \ref{fig:simpleTB}), obtained by fitting the bandwidth of the graphitic structure calculated using DFT.}
\label{tab:TBparam}
\end{table}

\begin{table}
\centering
\begin{tabular}{c c c c c c c}
\toprule
 & $J_{x1}$ & $J_{x2}$ & $J_{x1}'$ & $J_{x2}'$ & $J_{y1}$ & $J_{y2}$ \\
\midrule
C1 & 0.2000 & $-$0.1820 & $-$0.1820 & 0.2000 & $-$0.2067 & -0.2710 \\
C2 & 0.2100 & $-$0.1820 & $-$0.2200 & 0.2000 & $-$0.2067 & -0.2710 \\
D1 & 0.1400 & $-$0.1600 & $-$0.3300 & 0.0300 & $-$0.3044 & 0.1196 \\
D2 & 0.0300 & $-$0.1600 & $-$0.1600 & 0.0300 & $-$0.3044 & 0.1196 \\
\bottomrule
\end{tabular}
\caption{Tight-binding parameters used to create domain walls along the $x$ direction in Fig. \ref{fig:TBGnr}b}
\label{tab:TBparam2}
\end{table}

\begin{table}
\centering
\begin{tabular}{c c c c c c c}
\toprule
 & $J_{x1}$ & $J_{x2}$ & $J_{x1}'$ & $J_{x2}'$ & $J_{y1}$ & $J_{y2}$ \\
\midrule
A1 & 0.20 & $-$0.30 & $-$0.20 & 0.30 & $-$0.20 & $-$0.30 \\
A2 & 0.30 & $-$0.20 & $-$0.30 & 0.20 & $-$0.20 & $-$0.30 \\
B1 & 0.20 & $-$0.30 & $-$0.20 & 0.30 & $-$0.30 & 0.20 \\
B2 & 0.30 & $-$0.20 & $-$0.30 & 0.20 & $-$0.30 & 0.20 \\
C1 & 0.24 & $-$0.26 & $-$0.24 & 0.26 & $-$0.20 & $-$0.30 \\
C2 & 0.26 & $-$0.24 & $-$0.26 & 0.24 & $-$0.20 & $-$0.30 \\
D1 & 0.30 & $-$0.20 & $-$0.30 & 0.20 & $-$0.24 & $-$0.26 \\
D2 & 0.30 & $-$0.20 & $-$0.30 & 0.20 & $-$0.26 & $-$0.24 \\
\bottomrule
\end{tabular}
\caption{Tight-binding parameters for the artificial lattices used to create domain walls along both $x$ and $y$ directions in Fig. 4a-4c. Parameters in unlabeled regions can be obtained by mirror symmetry.}
\label{tab:TBparam3}
\end{table}

\begin{table}
\centering
\begin{tabular}{c c c c c c c}
\toprule
 State& $\Sigma_{ii}$ & $l$ & $n$ & $\Sigma_{ii}+|l|+2n$  \\
\midrule
Massless state & $-$1 & 0 & 0 & $-$1 \\
$s$-like massive state & 0 & 0 & 0 & 0 \\
$s$-like massive state & 0 & 0 & 0& 0 \\
$p$-like massive state & $-$1 & $-$1 & 0& 0 \\
$p$-like massive state & $-$1 & 1 & 0& 0 \\

\bottomrule
\end{tabular}
\caption{Number of degenerate states per corner and corresponding angular momentum ($l$) as analyzed in the main text.}
\label{tab:degeneracy}
\end{table}

